\documentclass{elsart}
\usepackage{graphicx,amssymb}
\usepackage{epsfig}
\begin{document}
\begin{frontmatter}

\title{Finite element computation of absorbing boundary conditions
for time-harmonic wave problems}

\author[Lami]{Denis Duhamel\corauthref{cor}},
\author[Lami]{Tien-Minh Nguyen}
\corauth[cor]{Corresponding author, duhamel@lami.enpc.fr.}

\address[Lami]{Universit\'e Paris-Est, UR Navier, \\
Ecole des Ponts ParisTech, \\
6 et 8 Avenue Blaise Pascal, \\
Cit\'{e} Descartes, Champs sur Marne,\\
77455 Marne la Vall\'{e}e, cedex 2, France\\
Tel: 33 1 64 15 37 28 \\
Fax: 33 1 64 15 37 41 \\
email{\nobreakspace}: duhamel@lami.enpc.fr}

\vspace{5cm}

Number of pages : 47

Number of figures : 17

\date{}
\newpage

\begin{abstract}
This paper proposes a new method, in the frequency domain, to define
absorbing boundary conditions for general two-dimensional problems.
The main feature of the method is that it can obtain boundary
conditions from the discretized equations without much knowledge of
the analytical behavior of the solutions and is thus very general.
It is based on the computation of waves in periodic structures and
needs the dynamic stiffness matrix of only one period in the medium
which can be obtained by standard finite element software. Boundary
conditions at various orders of accuracy can be obtained in a simple
way. This is then applied to study some examples for which
analytical or numerical results are available. Good agreements
between the present results and analytical solutions allow to check
the efficiency and the accuracy of the proposed method.
\end{abstract}

\begin{keyword}
Absorbing boundary conditions, waveguide, finite element, periodic medium.
\end{keyword}
\end{frontmatter}

\newpage
\section{Introduction}

Wave problems in unbounded media can occur in many applications in
mechanics and engineering such as in acoustics, solid mechanics,
electromagnetics, etc. It is well known that analytical solutions
for such problems are available only for some special cases. On the
contrary, numerical methods can be applied to many complex problems.
Physically, for problems in infinite domains, the energy is produced
by sources in the region to be analyzed and must escape to infinity.
For methods solving the problem on a bounded domain like the finite
element method, it introduces the difficulty of an artificial
boundary to get a bounded domain. This boundary must be such that
the energy crosses it without reflection and special conditions must
be specified at the artificial boundary to reproduce this phenomena.
Generally, these can be classified into local or global boundary
conditions. With a global condition all degrees of freedom (dofs) on
the boundary are coupled while a local condition connects only
neighboring dofs.

The first global method which has been used for solving such problems
was the boundary element method.
This method is well adapted for infinite domains
and is described in numerous classical textbooks like \cite{Bre1,Cro1,Cis1,Chen92,Bon1}.
It consists in solving an equation on the boundary of the domain only
and the radiation conditions are taken into account analytically.
It also reduces the dimension of the problem to a surface in 3D and
to a curve in 2D decreasing thus the size of the linear problem to solve.
However, the final problem involves full matrices which are also generally
non symmetrical.
It is also mainly limited to linear problems and to homogeneous domains
or otherwise one has to introduce special and complex techniques
to deal with non linear or non homogeneous situations.
There are also singularities in the integrals which need special attention
for the numerical integrations.
So this method is interesting and has been extensively used
but it can lead to heavy computations when the number of degrees of freedom increases.
More information on such techniques can be found in the historical
and review papers \cite{Che1,Bes2}.

In the other approaches, the computational domain is truncated at some distance
and boundary conditions are imposed at this artificial boundary.
These conditions at finite distance must simulate as closely as possible
the exact radiation condition at infinity.
An approach leading to a global boundary condition is the Dirichlet
to Neumann (DtN) mapping proposed by \cite{Keller89,Keller89b}
and in an earlier version by \cite{Hunt74,Hunt75}.
It consists in dividing the domain into a finite part containing the sources
and an infinite domain of simple shape.
The solution in the infinite domain is solved analytically, for example
by series expansions, and an exact impedance relation is obtained
on the boundary between the finite and infinite domains.
This relation links the variable and its normal derivative on the whole boundary.
The DtN mapping is thus non local and every node on the boundary is connected
to all other nodes. This gives a full matrix for the nodes of the boundary
which partially destroys the sparse matrix of the FEM and increases
substantially the computing resources needed to get the solution.
The solution has to be found in the exterior domain by analytical or numerical methods.
When the analytical solution can be found, it is generally under the form of
a series expansion. The number of terms in the expansion must be sufficient
for an accurate solution which can lead to heavy computations.
Developments of the method can be found in \cite{Harari91,Thompson99}.
An application to the case of wave scattering in plates is also found in \cite{Gal1}.

The other methods are local and the condition at a node involves
only neighboring nodes which make them less demanding in computing
resources and much easier to implement in a finite element code but
also less accurate. A first possibility of such approaches is the
use of infinite elements as proposed by
\cite{Bettes77,Bettes80,Bettes84,Bettes92,Astley00}. It consists in
developing special elements with a behavior at infinity reflecting
that of analytical solutions obtained for the same type of problems.
For wave problems, it involves complex-valued basis functions with
outwarding propagation wave-like behavior in the radial direction.
The elements were further developed by \cite{Burnett94} to
considered other coordinate systems such as expansions in prolate
coordinates. This method is interesting but the inclusion of
infinite elements requires the development of special elements and
these elements can depend on decay parameters which have to be
accurately chosen. A review of these methods has been proposed by
\cite{Gerdes00}.

In the perfectly matched layer proposed by
\cite{Berenger94,Berenger96}, originally for electromagnetic waves,
an exterior layer of finite thickness is introduced around the
bounded domain. The absorption in this domain is increasing as we
move towards the exterior such that outgoing waves are absorbed
before reaching the exterior truncation boundary. The number of
elements in the layer, its thickness, the variation of the
absorption properties have to be carefully chosen to optimize the
efficiency of the method. This efficiency is better for a layer with
a large thickness but this can lead to a significant increase in the
number of elements in the finite element model. Various developments
of the method can be found in \cite{Collino98} and its optimization
in \cite{Asv1}. Otherwise, various classes of absorbing boundary
conditions were also developed by \cite{Engquist77}. They consist in
the numerical approximation of differential operators on the
boundary. For instance, examples of the application of
Bayliss-Turkel conditions are presented by \cite{Str1}. However, the
more accurate boundary conditions involve high order derivatives on
the boundary which are difficult to implement in the finite element
method \cite{Pinsky92}. Finally, it was proved by \cite{Asv1,Gud1}
that, in fact, the perfectly matched layer and the absorbing
boundary conditions were closely connected. The Helmholtz equation
was also solved with these two boundary conditions by \cite{Hei1}
and these conditions were compared and optimized to minimize the
reflection. Other boundary conditions involving only second order
derivatives have also been proposed. They introduce auxiliary
variables and systems of equations on the boundary which lead to
high order boundary conditions, see \cite{Givoli04} for a review of
such non-reflecting boundary (NRBC) methods. They were mainly
developed for acoustic problems but in \cite{Krenk01} a local
boundary condition for elastic waves has been proposed. In
\cite{Gua1} an impedance boundary condition in new coordinates was
developed for the convected Helmholtz equation. For fluid dynamic
problems, \cite{Sto1} developed Lagrange multipliers for imposing
various absorbing boundary conditions for cases where the type and
the number of boundary conditions can change, for instance as the
flow changes from subsonic to supersonic regimes and its direction
varies with time. A general review of the methods described in the
precedent paragraphs for various dynamic, acoustic and wave
propagation problems can be found in
\cite{Givoli91,Mesquita05,Harari06,Thompson06}. Comparisons are also
made between the different methods.

In the present study, another local method is proposed. This method
works on discrete systems directly, in contrast with many existing
absorbing boundary conditions which are written on the continuous
differential equations and discretized after. The principle of the
method is to compute wave propagations in groups of elements near
the boundary from the dynamic stiffness matrix of these elements.
Then, a boundary condition is obtained for cancelling the reflected
waves. This condition is finally written as an impedance boundary
condition relating the force and displacement degrees of freedom on
the boundary. The approach is based on the waveguide theory proposed
by \cite{Mac1,Duh1,Duhamel06a} and is used to determine absorbing
boundary conditions at the truncation boundary of 2D periodic media.
Only information related to one period, obtained from any standard
FE software (the discrete stiffness and mass matrices and the nodal
coordinates) are required to formulate the method. The advantage is
that it can be applied to media with various complex behaviors.

This paper is outlined as follows. In section 2, the methodology for
determining absorbing boundary conditions for periodic media is
described. Then, a discussion for the application of the method to
general media is proposed. In section 3, a simple application is
proposed to show the results of the method in a case where detailed
computations can be done. In section 4, two examples of finite
element computations in acoustics and elastodynamics are presented.
They allow to check the efficiency and accuracy of the proposed
method for more complex cases. Finally, the paper is closed with
some conclusions.

\section{Absorbing boundary conditions}

\begin{figure}
  \begin{center}
    \epsfig{file=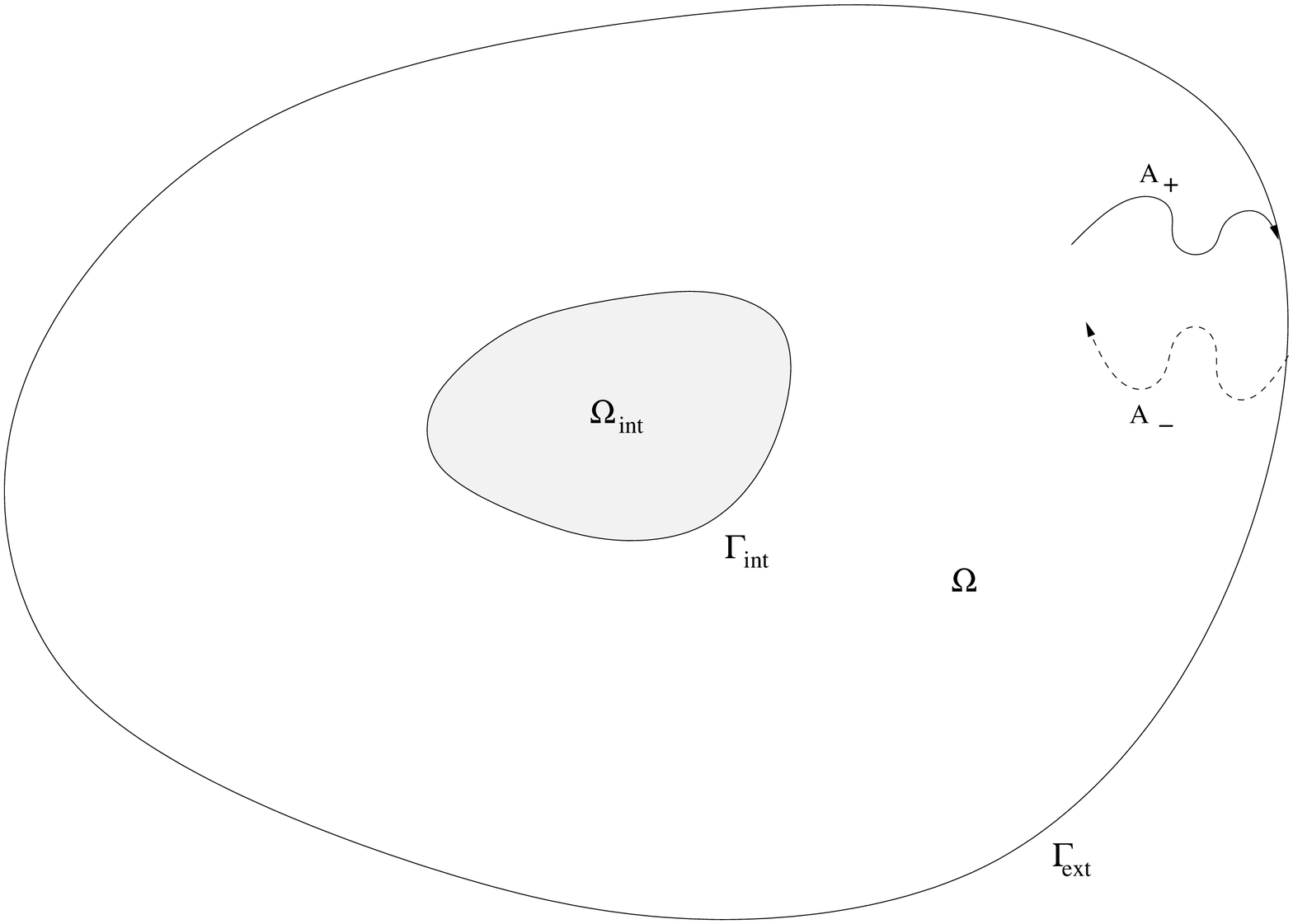,scale=.5}
    \caption{Computational domain}
    \label{fig01}
  \end{center}
\end{figure}

We suppose that we want to solve a mechanical problem on an infinite
domain exterior to the bounded domain $\Omega_{int}$ (see figure
\ref{fig01}). The infinite domain is approximated by the finite
domain $\Omega$ which is exterior to $\Omega_{int}$ and is limited
by the exterior boundary $\Gamma_{ext}$. We are looking for a
solution with radiating condition at infinity which means that the
solution should be outgoing near the boundary $\Gamma_{ext}$. Near
this exterior boundary the solution can be seen as composed of
incident waves denoted $A_+$ and reflected waves $A_-$. For a
perfectly absorbing boundary, one should have $A_-=0$. In fact this
condition is very difficult to implement in the numerical solutions
of such problems. Indeed, only the  global solution is easily
computed but the decomposition into incident and reflected waves is
difficult to obtain. The problem is thus to find an appropriate
boundary condition to impose on the exterior boundary to finally get
$A_-\approx 0$ on the solution. To be easily included in a finite
element model the searched boundary condition should be local and
the condition at a node of the boundary should involve only
neighboring nodes.

The approach proposed in this paper consists in studying this
problem by first considering the case of periodic media. For this
case, positive and negative waves and their amplitudes $A_+$ and
$A_-$ can be computed by the method presented below. Then an exact
boundary condition can be formulated for a half-plane boundary. It
is further shown how this condition can be approximated by a local
condition on the boundary. As homogeneous media are special cases of
periodic media, the method presented here applies also to
homogeneous media. Before considering the general case, a simple
example for the Klein Gordon equation will be presented.

\subsection{A simple example}

Consider first the stationary Klein Gordon equation given by
\begin{equation}
\frac{d^2 u}{dx^2} + (k^2-m^2)u = 0
\label{eqc01}
\end{equation}
where $u$ is the solution and $k$, $m$ are real parameters. This
equation is discretized with linear two nodes elements such that
\begin{equation}
u(\xi) = u_1N_1(\xi) + u_2N_2(\xi)
\end{equation}
where $N_1(\xi) = (1-\xi)/2$, $N_2(\xi)=(1+\xi)/2$, $u_1$ and $u_2$
are the values of the function at the two nodes of the element.
The discretization of the first and second parts of relation (\ref{eqc01})
leads, for an element of length $l$, to the matrices
\begin{equation}
\mathbf{k} = -\frac{1}{l}\left[
\begin{array}{cc}
1 & -1 \\
-1 & 1
\end{array}\right]\ \ \ \ \
\mathbf{m} = \frac{l}{6}\left[
\begin{array}{cc}
2\ & 1 \\
1\ & 2
\end{array}\right]
\end{equation}
and the dynamic stiffness matrix of one element is given by
\begin{equation}
\mathbf{d} = -\frac{1}{l}\left[
\begin{array}{cc}
1 & -1 \\
-1 & 1
\end{array}\right]
+ \frac{l}{6}(k^2-m^2)\left[
\begin{array}{cc}
2\ & 1 \\
1\ & 2
\end{array}\right]
\end{equation}
Waves of propagating constant $e^{\mu}$ are such that
\begin{eqnarray}
u_2 & = & e^{\mu}u_1 \\
f_2 + e^{\mu} f_1 & = & 0
\end{eqnarray}
leading in an element to
\begin{equation}
e^{2\mu}d_{12}+(d_{11}+d_{22})e^{\mu}+d_{21} = 0
\end{equation}
where $d_{ij}$ are the components of the matrix $\mathbf{d}$.
Taking into account the symmetries in the matrix $\mathbf{d}$,
this yields
\begin{equation}
e^{2\mu}+2\frac{d_{11}}{d_{12}}e^{\mu}+1 = 0
\end{equation}
whose solutions are
\begin{equation}
e^{\mu_{\pm}} = -\frac{d_{11}}{d_{12}}\pm \sqrt{(\frac{d_{11}}{d_{12}})^2-1} \\
\end{equation}
with
\begin{equation}
\frac{d_{11}}{d_{12}} = -\frac{1-l^2(k^2-m^2)/3}{1+l^2(k^2-m^2)/6}
\end{equation}
For $l^2(k^2-m^2) << 1$, one gets, at first order,
\begin{equation}
e^{\mu_{\pm}} \approx 1 \pm il\sqrt{k^2-m^2}
\end{equation}
meaning
\begin{equation}
\mu_{\pm} \approx \pm il\sqrt{k^2-m^2}
\end{equation}
There are thus two waves in an element, such that
\begin{equation}\left[
\begin{array}{c}
u_+ \\
f_+
\end{array}\right]
=\left[
\begin{array}{c}
1 \\
d_{11} + e^{\mu_+}d_{12}
\end{array}\right]
\ \ and \ \
\left[
\begin{array}{c}
u_- \\
f_-
\end{array}\right]
=\left[
\begin{array}{c}
1 \\
d_{11} + e^{\mu_-}d_{12}
\end{array}\right]
\end{equation}
and the general solution is given by
\begin{equation}
\left[
\begin{array}{c}
u \\
f
\end{array}\right]=a_+
\left[
\begin{array}{c}
u_+ \\
f_+
\end{array}\right]
+ a_-
\left[
\begin{array}{c}
u_- \\
f_-
\end{array}\right]
\end{equation}
The condition for only outgoing waves is thus $a_-=0$ on the right
boundary  and $a_+=0$ on the left boundary leading respectively to
the conditions
\begin{eqnarray}
f/u & = & f_+/u_+ \ \ on\ the\ right\nonumber \\
f/u & = & f_-/u_- \ \ on\ the\ left
\end{eqnarray}
The condition in the first case is
\begin{eqnarray}
f/u & = & d_{11} + e^{\mu_+}d_{12} \nonumber \\
& = & d_{11} + \left(-\frac{d_{11}}{d_{12}} +\sqrt{(\frac{d_{11}}{d_{12}})^2-1}\right)d_{12} \nonumber \\
& = & \sqrt{d_{11}^2 - d_{12}^2} \nonumber \\
& = & \sqrt{-(k^2-m^2) + \frac{1}{12}((k^2-m^2)l)^2} \nonumber \\
& = & i\sqrt{k^2-m^2}\sqrt{1 - \frac{1}{12}(k^2-m^2)l^2} \label{eq01a} \\
& \approx & i\sqrt{k^2-m^2}
\end{eqnarray}
In the second case, one gets
\begin{equation}
f/u = d_{11} + e^{\mu_-}d_{12} \approx -i\sqrt{k^2-m^2}
\end{equation}

We recognize approximations of the classical absorbing boundary
conditions which have been obtained here directly from the
discretized equations. Compared to the classical boundary condition
on the right (and exact in this case) $f/u = i\sqrt{k^2-m^2}$, the relative error
is $\sqrt{1 - \frac{1}{12}(k^2-m^2)l^2}$ which depends mainly on the
size of the element relatively to the wavelength. The present
boundary condition has been obtained entirely from the discrete
matrices without any knowledge of the analytical solution of the
problem.

To estimate the reflection coefficient created by such a boundary condition
consider an incident wave $Ae^{i\sqrt{k^2-m^2}x}$ on the boundary.
A reflected wave $RAe^{-i\sqrt{k^2-m^2}x}$ is created. 
The total solution and its associated force are given by
\begin{eqnarray}
u(x) & = & Ae^{i\sqrt{k^2-m^2}x} + RAe^{-i\sqrt{k^2-m^2}x} \nonumber \\
f(x) & = & Ai\sqrt{k^2-m^2}e^{i\sqrt{k^2-m^2}x} - RAi\sqrt{k^2-m^2}e^{-i\sqrt{k^2-m^2}x}
\end{eqnarray}
Writing the boundary condition (\ref{eq01a}), for instance by taking the boundary at $x=0$, yields
\begin{equation}
\frac{f(0)}{u(0)} = \frac{1-R}{1+R} = \sqrt{1 - \frac{1}{12}(k^2-m^2)l^2}
\end{equation}
So the reflection coefficient is finally given by
\begin{eqnarray}
R & = & \frac{1-\sqrt{1 - \frac{1}{12}(k^2-m^2)l^2}}{1+\sqrt{1 - \frac{1}{12}(k^2-m^2)l^2}} \nonumber \\
& \approx & \frac{1}{48}(k^2-m^2)l^2
\end{eqnarray}
This coefficient is low and of second order when $\sqrt{k^2-m^2}l << 1$.

\subsection{General impedance boundary condition}

\begin{figure}
  \begin{center}
    \epsfig{file=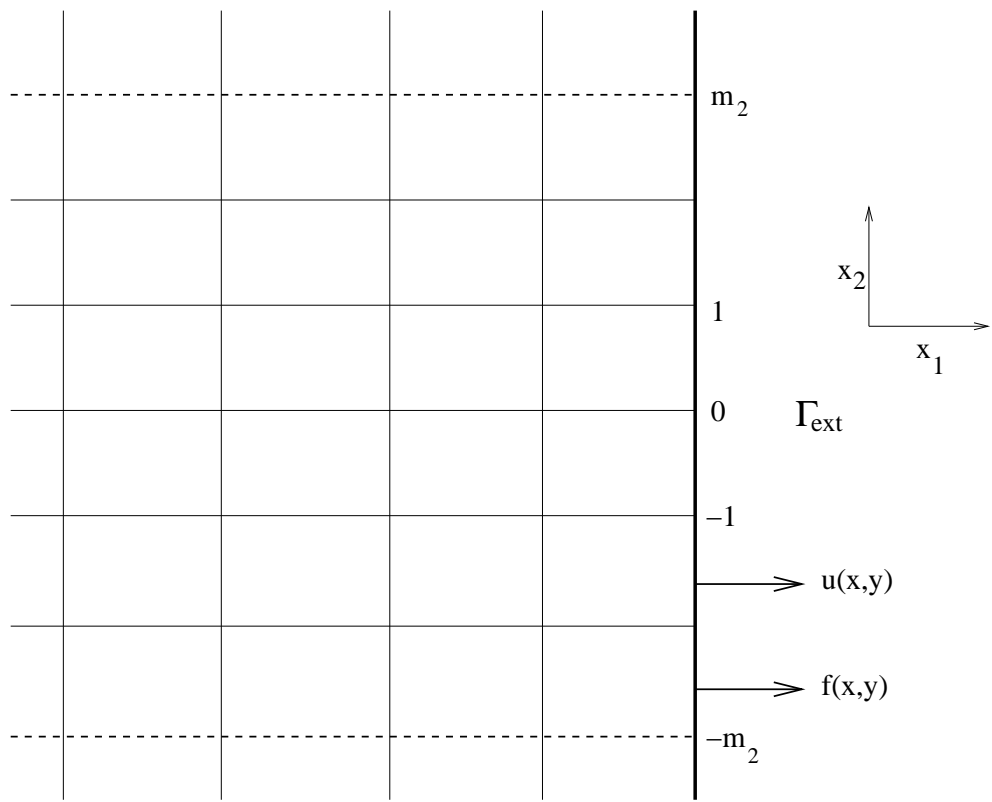,scale=.7}
    \caption{Periodic medium near the exterior boundary.}
    \label{fig04}
  \end{center}
\end{figure}

In this section we present the general outline of the method before
starting a more rigorous developement in the following section. 
So, to extend the precedent example to more general cases, consider a
vector function $\mathbf{u}(x,y)$ and a force vector 
$\mathbf{f}(x,y)$ acting on a line parallel to the $y$ axis as in
figure \ref{fig04}. They can be decomposed by a Fourier transform as
\begin{eqnarray}
\mathbf{u}(x,y) = \frac{1}{(2\pi)^2}\int_{-\infty}^{+\infty}
\int_{-\infty}^{+\infty} \mathbf{u}(k_x,k_y)e^{i(k_xx+k_yy)}dk_xdk_y
\nonumber \\
\mathbf{f}(x,y) = \frac{1}{(2\pi)^2}\int_{-\infty}^{+\infty}
\int_{-\infty}^{+\infty} \mathbf{f}(k_x,k_y)e^{i(k_xx+k_yy)}dk_xdk_y
\end{eqnarray}
Let us suppose that $\mathbf{u}$ is solution of a linear operator
\begin{equation}
L(\mathbf{u}) = \sum_{n=0}^{n=N}\sum_{\alpha_1+\alpha_2=n}
\mathbf{a}_{\alpha_1\alpha_2} \frac{\partial^n \mathbf{u}}{\partial
x^{\alpha_1}
\partial y^{\alpha_2}} = 0
\end{equation}
In the Fourier domain, this relation yields
\begin{equation}
\left(\sum_{n=0}^{n=N}\sum_{\alpha_1+\alpha_2=n}
\mathbf{a}_{\alpha_1\alpha_2} (ik_x)^{\alpha_1}(ik_y)^{
\alpha_2}\right)\mathbf{u}(k_x,k_y) = 0
\end{equation}
For a given value of $k_y$, the precedent relation has non zero
solutions for $k_x$ such that the determinant
\begin{equation}
\left|\sum_{n=0}^{n=N}\sum_{\alpha_1+\alpha_2=n}
\mathbf{a}_{\alpha_1\alpha_2} (ik_x)^{\alpha_1}(ik_y)^{ \alpha_2}\right| = 0
\end{equation}
Let us denote by $k_{j}^+$ the $n_+$ positive solutions such that
$Re(k_{j}^+) < 0$ or $Re(k_{j}^+) = 0$ and the energy flux is
directed towards positive values of x. We denote by $k_{j}^-$ the
other solutions. We have the decomposition
\begin{equation}
\mathbf{u}(x,k_y) = \sum_{j=1}^{j=n_+}a_j^+\mathbf{u}_j^+e^{ik^+_{j}x}+
\sum_{j=1}^{j=n_-}a_j^-\mathbf{u}_j^-e^{ik^-_{j}x}
\end{equation}
In the same way, for the force components
\begin{equation}
\mathbf{f}(x,k_y) = \sum_{j=1}^{j=n_+}a_j^+\mathbf{f}_j^+e^{ik^+_{j}x}+
\sum_{j=1}^{j=n_-}a_j^-\mathbf{f}_j^-e^{ik^-_{j}x}
\end{equation}
where $\mathbf{f}_j^+$ and $\mathbf{f}_j^-$ are the force components
respectively associated to $\mathbf{u}_j^+$ and $\mathbf{u}_j^-$.
If the boundary is such that only positive waves exists at
proximity, one has
\begin{eqnarray}
\mathbf{u}(0,k_y) & = & \sum_{j=1}^{j=n_+}a_j^+\mathbf{u}_j^+ = \mathbf{U}^+\mathbf{a}^+\nonumber \\
\mathbf{f}(0,k_y) & = & \sum_{j=1}^{j=n_+}a_j^+\mathbf{f}_j^+ =
\mathbf{F}^+\mathbf{a}^+
\end{eqnarray}
where $\mathbf{U}^+$ and $\mathbf{F}^+$ are the
matrices whose columns are respectively $\mathbf{u}_j^+$ and
$\mathbf{f}_j^+$. Eliminating the $\mathbf{a}^+$ coefficients, one
gets
\begin{eqnarray}
\mathbf{f}(0,k_y) & = & \mathbf{F}^+(\mathbf{U}^+)^{-1}\mathbf{u}(0,k_y) \nonumber \\
\mathbf{f}(0,y) & = & (\mathbf{G}*\mathbf{u})(0,y)
\end{eqnarray}
with
\begin{equation}
\mathbf{G}(k_y) =\mathbf{F}^+(\mathbf{U}^+)^{-1}(k_y)
\end{equation}
and $*$ means the convolution.

In the following this boundary condition will be computed directly from the discrete equations
for general linear media.

\subsection{Solution in a periodic medium}

\begin{figure}
  \begin{center}
    \epsfig{file=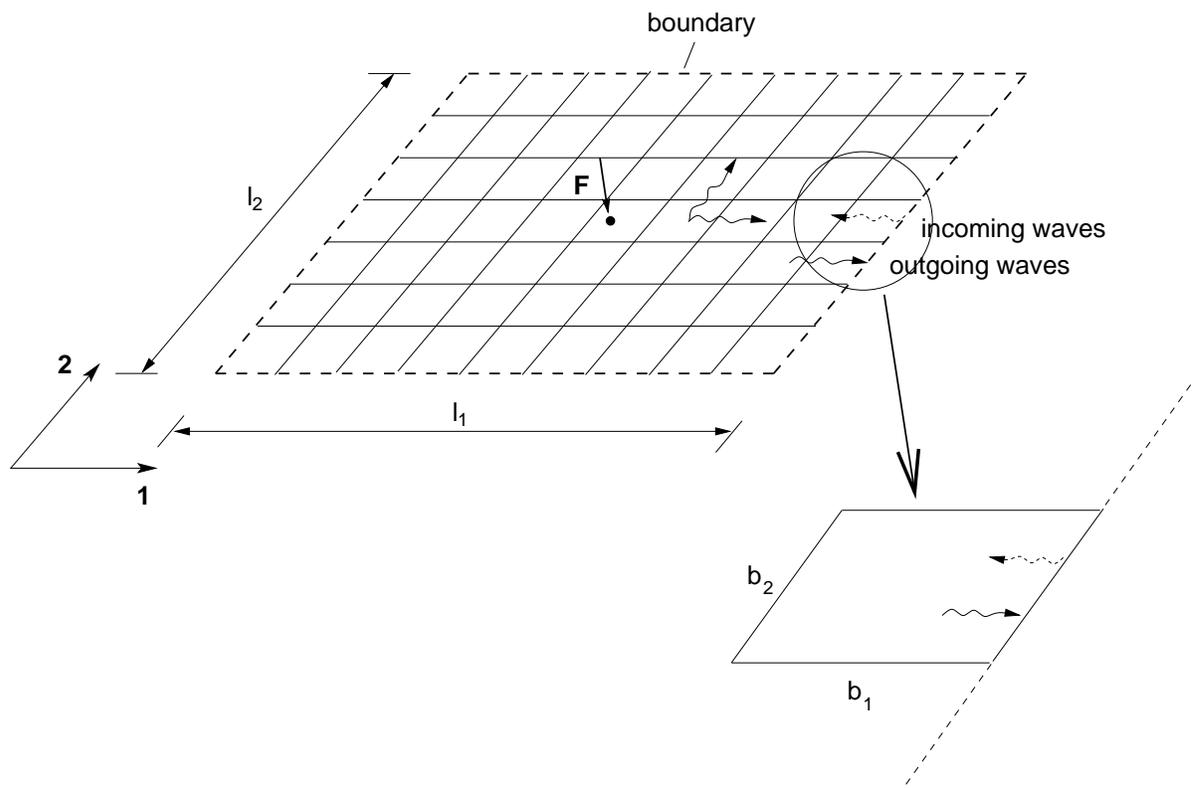,scale=.7}
    \caption{Periodic medium}
    \label{fig02}
  \end{center}
\end{figure}

Consider an infinite two dimensional periodic medium, as shown in
figure \ref{fig02}. The elementary period is limited by the domain
$(x_1,x_2) \in [0, b_1] \times [0, b_2]$.
A function $U(x_1,x_2)$ defined on the two-dimensional periodic medium
can be decomposed as an integral of pseudo periodic functions
\begin{equation}
U(x_1,x_2) = \int_{-\frac{\pi}{b_2}}^{\frac{\pi}{b_2}} e^{ikx_2}\hat{U}(x_1,k,x_2) dk
\label{eq00}
\end{equation}
where $\hat{U}(x_1,k,x_2)$ is a periodic function in $x_2$ with period $b_2$.
From the Fourier transform $\hat{U}(x_1,k)$ of $U(x_1,x_2)$, one has
\begin{eqnarray}
\hat{U}(x_1,k,x_2)
& = & \frac{1}{2\pi}\sum_{m_2=-\infty}^{+\infty}\hat{U}(x_1,k+2\pi\frac{m_2}{b_2})
e^{i2\pi \frac{m_2}{b_2}x_2} \nonumber \\
& = & \frac{1}{2\pi}\sum_{m_2=-\infty}^{+\infty}e^{i2\pi \frac{m_2}{b_2}x_2}
\int_{-\infty}^{+\infty}e^{-i(k+2\pi \frac{m_2}{b_2})x}U(x_1,x)dx \nonumber \\
& = & \frac{1}{2\pi}\int_{-\infty}^{+\infty}e^{-ikx}U(x_1,x)
\sum_{m_2=-\infty}^{+\infty}e^{i2\pi \frac{m_2}{b_2}(x_2-x)} dx \nonumber \\
& = & \frac{b_2}{2\pi}\int_{-\infty}^{+\infty}\sum_{m_2=-\infty}^{+\infty}\delta(x_2-x-m_2b_2)
e^{-ikx}U(x_1,x)dx \nonumber \\
& = & \frac{b_2}{2\pi}\sum_{m_2=-\infty}^{+\infty}e^{-ik(x_2+m_2b_2)}U(x_1,x_2+m_2b_2)
\label{eq01}
\end{eqnarray}
This gives the relation inverse of (\ref{eq00}).
From relation (\ref{eq00}), one sees that the behavior in $x_2$
of the general solution can be obtained from functions as
$e^{ikx_2}\hat{U}(x_1,k,x_2)$ with $\hat{U}(x_1,k,x_2)$ periodic in $x_2$.
Along direction $1$, we use a decomposition in Bloch waves as it is usual
in periodic media.
Finally, the general solution can be obtained from functions
$u(x_1,k,x_2)=e^{ikx_2}\hat{U}(x_1,k,x_2)$
such that:
\begin{eqnarray}
  u(x_1,k,x_2+m_2b_2) &=& e^{ikm_2b_2}u(x_1,k,x_2)\\
  u(x_1+m_1b_1,k,x_2) &=& e^{im_1\mu}u(x_1,k,x_2)
\label{eq02}
\end{eqnarray}

where $m_1$ and $m_2$ are integers, $k \in \mathbb{R}\cap\left
  [\displaystyle-\frac{\pi}{b_2}, \frac{\pi}{b_2}\right ]$ and
$\mu\in\mathbb{C}$.

The discrete dynamic equation of a cell (an elementary period) obtained
from a FE model at a frequency $\omega$ and for the time dependence
$e^{-i\omega t}$ is given by:
\begin{equation}
  (\mathbf{K} - i\omega\mathbf{C} - \omega^2\mathbf{M})\mathbf{q} =
  \mathbf{f}
\label{eq03}
\end{equation}

where $\mathbf{K}$, $\mathbf{M}$ and $\mathbf{C}$ are the
stiffness, mass and damping matrices, respectively, $\mathbf{f}$ is the
loading vector and $\mathbf{q}$ the vector of the degrees of
freedom (dofs). Introducing the dynamic stiffness matrix
$\widetilde{\mathbf{D}} = \mathbf{K} - i\omega\mathbf{C} -
\omega^2\mathbf{M}$, decomposing the dofs into boundary $(B)$ and interior
$(I)$ dofs, and assuming that there are no external forces on the
interior nodes, result in the following equation:
\begin{equation}
  \left [
    \begin{array}{cc}
      \widetilde{\mathbf{D}}_{BB} & \widetilde{\mathbf{D}}_{BI}\\
      \widetilde{\mathbf{D}}_{IB} & \widetilde{\mathbf{D}}_{II}
    \end{array}
  \right ]
  \left [
    \begin{array}{c}
      \mathbf{q}_{B}\\
      \mathbf{q}_{I}
    \end{array}
  \right ]
  =
  \left [
    \begin{array}{c}
      \mathbf{f}_{B}\\
      \mathbf{0}
    \end{array}
  \right ]
  \label{eq04}
\end{equation}

The interior dofs can be eliminated using the second row of
equation (\ref{eq04}), which results in
\begin{equation}
  \mathbf{q}_I =
  -\widetilde{\mathbf{D}}_{II}^{-1}\widetilde{\mathbf{D}}_{IB}\mathbf{q}_B
\label{eq05}
\end{equation}

The first row of equation (\ref{eq04}) becomes
\begin{equation}
  \mathbf{f}_B = \left
    (\widetilde{\mathbf{D}}_{BB}-\widetilde{\mathbf{D}}_{BI}\widetilde{\mathbf{D}}_{II}^{-1}\widetilde{\mathbf{D}}_{IB}\right )\mathbf{q}_B
\label{eq06}
\end{equation}

which can be written as
\begin{equation}
  \mathbf{f} = \mathbf{D}\mathbf{q}
  \label{eq07}
\end{equation}

It should be noted that only boundary dofs are considered in the following.

\begin{figure}
  \begin{center}
    \epsfig{file=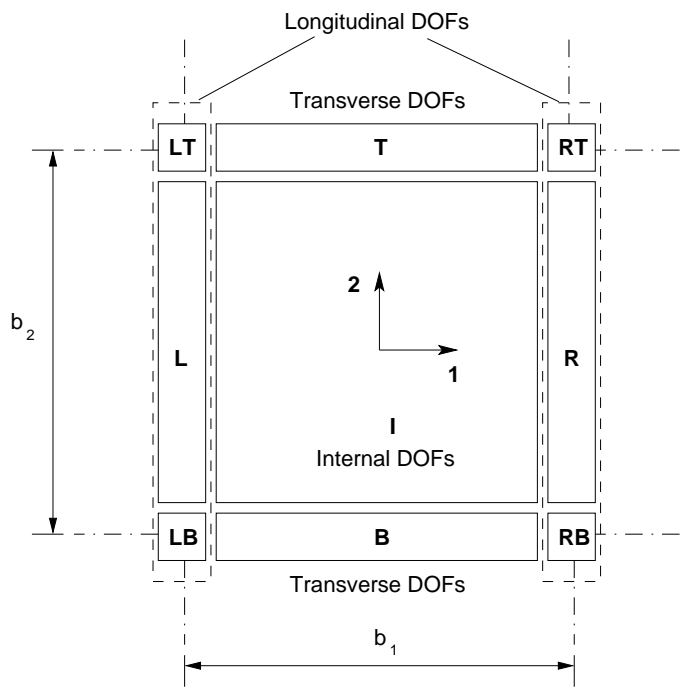,scale=.7}
    \caption{A cell in the periodic medium}
    \label{fig03}
  \end{center}
\end{figure}

The periodic cell is assumed to be meshed with an equal number of nodes on
their opposite sides. The boundary dofs are decomposed into left $(L)$,
right $(R)$, bottom $(B)$, top $(T)$ dofs and associated corners $(LB)$,
$(RB)$, $(LT)$ and $(RT)$ as shown in figure \ref{fig03}. The
longitudinal dofs vector is defined as $\mathbf{q}_l = {}^t\left
  [{}^t\mathbf{q}_L \quad
  {}^t\mathbf{q}_R \quad {}^t\mathbf{q}_{LB} \quad {}^t\mathbf{q}_{RB}
  \quad {}^t\mathbf{q}_{RT} \quad {}^t\mathbf{q}_{LT}\right ]$. Thus,
equation (\ref{eq07}) is rewritten as
\begin{equation}
  \left [
    \begin{array}{ccc}
      \mathbf{D}_{ll} & \mathbf{D}_{lB} & \mathbf{D}_{lT}\\
      \mathbf{D}_{Bl} & \mathbf{D}_{BB} & \mathbf{D}_{BT}\\
      \mathbf{D}_{Tl} & \mathbf{D}_{TB} & \mathbf{D}_{TT}
    \end{array}
  \right ]
  \left [
    \begin{array}{c}
      \mathbf{q}_l\\
      \mathbf{q}_B\\
      \mathbf{q}_T
    \end{array}
  \right ]
  =
  \left [
    \begin{array}{c}
      \mathbf{f}_l\\
      \mathbf{f}_B\\
      \mathbf{f}_T
    \end{array}
  \right ]
  \label{eq08}
\end{equation}

Using the pseudo periodic condition (\ref{eq02}) and the
effort equilibrium at the bottom side of the cell, relations between the
transverse dofs are given by
\begin{equation}
  \begin{array}{rcl}
    \mathbf{q}_T &=& e^{ikb_2}\mathbf{q}_B\\
    \mathbf{f}_B+e^{-ikb_2}\mathbf{f}_T &=& 0
  \end{array}
  \label{eq09}
\end{equation}

Multiplying the third row of equation (\ref{eq08}) with
$e^{-ikb_2}$, taking the sum of the second and third rows of
equation(\ref{eq08}), using conditions
(\ref{eq09}), lead to
\begin{equation}
  \left (\mathbf{D}_{Bl} + e^{-ikb_2}\mathbf{D}_{Tl}\right )\mathbf{q}_l +
  \left (\mathbf{D}_{BB} + \mathbf{D}_{TT} + e^{-ikb_2}\mathbf{D}_{TB} +
    e^{ikb_2}\mathbf{D}_{BT}\right )\mathbf{q}_B = 0
\label{eq10}
\end{equation}

so
\begin{equation}
  \mathbf{q}_B = -\left (\mathbf{D}_{BB} + \mathbf{D}_{TT} + e^{-ikb_2}\mathbf{D}_{TB} +
    e^{ikb_2}\mathbf{D}_{BT}\right )^{-1}\left (\mathbf{D}_{Bl} +
    e^{-ikb_2}\mathbf{D}_{Tl}\right )\mathbf{q}_l
  \label{eq11}
\end{equation}

Using (\ref{eq09}) and (\ref{eq11}),
the first row of equation (\ref{eq08}) becomes
\begin{eqnarray}
  \mathbf{f}_l &=& \left [\mathbf{D}_{ll}-\left (\mathbf{D}_{lB} +
    e^{ikb_2}\mathbf{D}_{lT}\right )\left (\mathbf{D}_{BB} +
      \mathbf{D}_{TT} + e^{-ikb_2}\mathbf{D}_{TB} +
      e^{ikb_2}\mathbf{D}_{BT}\right )^{-1}\right.\nonumber\\
  &&\left .\times\left (\mathbf{D}_{Bl} +
    e^{-ikb_2}\mathbf{D}_{Tl}\right )\right ]\mathbf{q}_l
    \label{eq12}
\end{eqnarray}

which can be written as
\begin{equation}
  \mathbf{f}_l = \mathbf{D}_l\mathbf{q}_l
  \label{eq13}
\end{equation}

Using the pseudo periodic conditions (\ref{eq02}) also lead
to the following relations between longitudinal dofs
\begin{equation}
  \begin{array}{lcl}
    \mathbf{q}_R & = & e^{i\mu}\mathbf{q}_L\\
    \mathbf{q}_{RB} & = & e^{i\mu}\mathbf{q}_{LB}\\
    \mathbf{q}_{RT} & = & e^{i(\mu+kb_2)}\mathbf{q}_{LB}\\
    \mathbf{q}_{LT} & = & e^{ikb_2}\mathbf{q}_{LB}
  \end{array}
  \label{eq14}
\end{equation}

From the pseudo periodic conditions (\ref{eq14}), it can
be seen that all components of the vector $\mathbf{q}_l$ depend on the
set of dofs defined by $\mathbf{q}_r = {}^t\left [{}^t\mathbf{q}_L\quad
{}^t\mathbf{q}_{LB}\right ]$. This can be expressed as
\begin{equation}
  \mathbf{q}_l = \left (\mathbf{W}_0 + e^{i\mu}\mathbf{W}_1\right
  )\mathbf{q}_r
  \label{eq15}
\end{equation}

where the matrices $\mathbf{W}_0$ and $\mathbf{W}_1$ depend on the
wavenumber $k$ and are given by
\begin{equation}
  \mathbf{W}_0 =
  \left [
    \begin{array}{cc}
      \mathbf{I} & \mathbf{O}\\
      \mathbf{O} & \mathbf{O}\\
      \mathbf{O} & \mathbf{I}\\
      \mathbf{O} & \mathbf{O}\\
      \mathbf{O} & \mathbf{O}\\
      \mathbf{O} & e^{ikb_2}\mathbf{I}
    \end{array}
  \right ]
  \qquad
  \mathbf{W}_1 =
  \left [
    \begin{array}{cc}
      \mathbf{O} & \mathbf{O}\\
      \mathbf{I} & \mathbf{O}\\
      \mathbf{O} & \mathbf{O}\\
      \mathbf{O} & \mathbf{I}\\
      \mathbf{O} & e^{ikb_2}\mathbf{I}\\
      \mathbf{O} & \mathbf{O}
    \end{array}
  \right ]
  \label{eq16}
\end{equation}

The equilibrium conditions between adjacent cells are given by
\begin{equation}
  \begin{array}{rcl}
    e^{i\mu}\mathbf{f}_L + \mathbf{f}_R &=& 0\\
    e^{i\mu}\mathbf{f}_{LB} + \mathbf{f}_{RB} +
    e^{i(\mu-kb_2)}\mathbf{f}_{LT} + e^{-ikb_2}\mathbf{f}_{RT} &=& 0
  \end{array}
\label{eq17}
\end{equation}

that can be written as
\begin{equation}
  \left (e^{i\mu}\mathbf{W}_0^* + \mathbf{W}_1^*\right )\mathbf{f}_l = 0
  \label{eq18}
\end{equation}

where $(.)^*$ denotes the operator of complex conjugate and
transpose.

Combining (\ref{eq13}), (\ref{eq15}) and
(\ref{eq18}), lead to
\begin{equation}
  \left (e^{i\mu}\mathbf{W}_0^* + \mathbf{W}_1^*\right
  )\mathbf{D}_l\left (\mathbf{W}_0 + e^{i\mu}\mathbf{W}_1\right
  )\mathbf{q}_r = 0
\label{eq19}
\end{equation}

that can be written as
\begin{equation}
  \left (\mathbf{A}_0 + e^{i\mu}(\mathbf{A}_1 + \mathbf{A}_2)
    +e^{2i\mu}\mathbf{A}_3\right )\mathbf{q}_r = 0
  \label{eq20}
\end{equation}

where
\begin{equation}
  \begin{array}{rcl}
    \mathbf{A}_0 & = & \mathbf{W}^*_1\mathbf{D}_l\mathbf{W}_0\\
    \mathbf{A}_1 & = & \mathbf{W}^*_0\mathbf{D}_l\mathbf{W}_0\\
    \mathbf{A}_2 & = & \mathbf{W}^*_1\mathbf{D}_l\mathbf{W}_1\\
    \mathbf{A}_3 & = & \mathbf{W}^*_0\mathbf{D}_l\mathbf{W}_1
  \end{array}
\label{eq21}
\end{equation}

The eigenvalue $e^{i\mu}$ and the eigenvector $\mathbf{q}_r$ are
thus solutions of a quadratic eigenvalue problem. It is convenient to
transform the problem (\ref{eq20}) into another linear
eigenvalue problem as
\begin{equation}
  e^{i\mu}\left [
    \begin{array}{cc}
      \mathbf{A}_3 & \mathbf{O}\\
      \mathbf{O} & \mathbf{A}_3
    \end{array}
  \right ]
  \left [
    \begin{array}{c}
      \mathbf{q}_r\\
      \widetilde{\mathbf{q}}_r
    \end{array}
  \right ]
  =
  \left [
    \begin{array}{cc}
      \mathbf{O} & \mathbf{A}_3\\
      -\mathbf{A}_0 & -(\mathbf{A}_1+\mathbf{A}_2)
    \end{array}
  \right ]
  \left [
    \begin{array}{c}
      \mathbf{q}_r\\
      \widetilde{\mathbf{q}}_r
    \end{array}
  \right ]
  \label{eq22}
\end{equation}

with $\widetilde{\mathbf{q}}_r = e^{i\mu}\mathbf{q}_r$.

From equations (\ref{eq12}) and (\ref{eq13}), one can
notice that
\begin{equation}
    {}^t\mathbf{D}_l(k) = \mathbf{D}_l(-k)
\label{eq23}
\end{equation}
Moreover, from (\ref{eq16}), we have
\begin{equation}
\mathbf{W}_j^*(k) = {}^t\overline{\mathbf{W}}_j(k) =
{}^t\mathbf{W}_j(-k) \qquad \mathrm{for} \quad j=0, 1
\label{eq24}
\end{equation}
and from (\ref{eq21})
\begin{eqnarray}
{}^t\mathbf{A}_0(k) & = & \mathbf{A}_3(-k) \nonumber \\
{}^t\mathbf{A}_1(k) & = & \mathbf{A}_1(-k) \nonumber \\
{}^t\mathbf{A}_2(k) & = & \mathbf{A}_2(-k) \nonumber \\
{}^t\mathbf{A}_3(k) & = & \mathbf{A}_0(-k)
\label{eq25}
\end{eqnarray}

It can be easily shown by taking the determinant of the matrix in relation
(\ref{eq20}) that if $e^{i\mu_j}$ is an eigenvalue for
the wavenumber $k$, $e^{-i\mu_j}$ is also an eigenvalue for the
wavenumber $-k$. These represent a pair of positive and
negative-going waves, respectively. The $2n$ eigensolutions of equation
(\ref{eq22}) can be split into two sets of $n^+$ and
$n^-$ eigensolutions with $2n = n^++n^-$, which are denoted by
$\left(e^{i\mu_j^+},\mathbf{q}_j^+\right)$ and
$\left(e^{i\mu_j^-},\mathbf{q}_j^-\right)$ respectively, with the first set
such that $\left|e^{i\mu_j^+}\right|\leq 1$. In the case
$\left|e^{i\mu_j^+}\right|=1$,
the first set of positive-going waves must contain waves
propagating in the positive direction such that
$\mathrm{Re}\left\{i\omega\mathbf{q}^H_j\mathbf{f}_j^r\right\}>0$ where
$\mathbf{f}_j^r$ is the reduced set of boundary force dofs of
left cells on right cells and is given by
\begin{equation}
  \mathbf{f}_j^r =
  \left [
    \begin{array}{c}
      \mathbf{f}_L\\
      \mathbf{f}_{LB}+e^{-ikb_2}\mathbf{f}_{LT}
    \end{array}
  \right ] =
  \mathbf{W}_0^*\mathbf{f}_j^l =
  \mathbf{W}_0^*\mathbf{D}_l\left (\mathbf{W}_0 + e^{i\mu_j}\mathbf{W}_1\right
  )\mathbf{q}_j
\label{eq26}
\end{equation}

In the second set of negative-going waves, the eigenvalues
$e^{i\mu_j^-}$ are associated with waves such that
$\mathrm{Re}\left\{i\omega\mathbf{q}^H_j\mathbf{f}_j^r\right\}<0$.

With the eigenvector $\mathbf{q}_j$ and the force component of relation (\ref{eq26}),
we introduce the state vector
\begin{equation}
\mathbf{x}_j(k) =
\left[
\begin{array}{c}
\mathbf{q}_j(k) \\
\mathbf{f}_j^r(k)
\end{array} \right] =
\left[
\begin{array}{c}
\mathbf{q}_j(k) \\
(\mathbf{A}_1(k)+e^{i\mu_j(k)}\mathbf{A}_3(k))\mathbf{q}_j(k)
\end{array} \right]
\end{equation}
In this relation $\mathbf{q}_j(k)$ is the eigenvector associated to $e^{i\mu_j(k)}$.
One can also introduce
\begin{equation}
\mathbf{y}_j(-k) =
\left[
\begin{array}{cc}
{}^t\mathbf{p}_j(-k)(\mathbf{A}_2(k)+e^{i\mu_j(k)}\mathbf{A}_3(k)) \ \ & \ \ {}^t\mathbf{p}_j(-k)
\end{array} \right]
\label{eq27}
\end{equation}
In this relation $\mathbf{p}_j(-k)$ is the eigenvector associated to $e^{-i\mu_j(k)}$
since we have seen that $e^{-i\mu_j(k)}$ is also an eigenvalue of (\ref{eq20}) for
the wavenumber $-k$.
From relation (\ref{eq20}) written for the eigenvector $\mathbf{q}_j(k)$, multiplying
this relation by $e^{-i\mu_j(k)}$ and then on the left by ${}^t\mathbf{p}_i(-k)$, one gets
\begin{equation}
{}^t\mathbf{p}_i(-k)\left (e^{-i\mu_j(k)}\mathbf{A}_0(k) + (\mathbf{A}_1(k) + \mathbf{A}_2(k))
+e^{i\mu_j(k)}\mathbf{A}_3(k)\right )\mathbf{q}_j(k) = 0
\end{equation}
In the same way, writing relation (\ref{eq20}) for the eigenvector $\mathbf{p}_i(-k)$,
taking the transpose of the relation, using relations (\ref{eq25}) and multiplying
on the right by $\mathbf{q}_j(k)$, leads,
after a global multiplication by $e^{i\mu_i(k)}$, to the following relation
\begin{equation}
{}^t\mathbf{p}_i(-k)\left (e^{i\mu_i(k)}\mathbf{A}_3(k) + (\mathbf{A}_1(k) + \mathbf{A}_2(k))
    +e^{-i\mu_i(k)}\mathbf{A}_0(k)\right ) \mathbf{q}_j(k) =  0
\end{equation}
The difference between the two precedent relations yields
\begin{equation}
(e^{i\mu_i(k)}-e^{i\mu_j(k)})
{}^t\mathbf{p}_i(-k)\left (\mathbf{A}_3(k)
    -e^{-i\mu_i(k)}e^{-i\mu_j(k)}\mathbf{A}_0(k)\right) \mathbf{q}_j(k) =  0
\end{equation}
In the case $e^{i\mu_i(k)}\neq e^{i\mu_j(k)}$, we get
\begin{equation}
{}^t\mathbf{p}_i(-k)\left (e^{i\mu_i(k)}\mathbf{A}_3(k)
    -e^{-i\mu_j(k)}\mathbf{A}_0(k)\right) \mathbf{q}_j(k) = 0
\label{eq28}
\end{equation}

Now it is possible to compute the product $\mathbf{y}_i(-k).\mathbf{x}_j(k)$ by
\begin{eqnarray}
\mathbf{y}_i(-k).\mathbf{x}_j(k)
& = & {}^t\mathbf{p}_i(-k)(\mathbf{A}_2(k)+e^{i\mu_i(k)}\mathbf{A}_3(k))\mathbf{q}_j(k) \nonumber \\
& & +{}^t\mathbf{p}_i(-k)(\mathbf{A}_1(k)+e^{i\mu_j(k)}\mathbf{A}_3(k))\mathbf{q}_j(k) \nonumber \\
& = & {}^t\mathbf{p}_i(-k)(\mathbf{A}_2(k)+e^{i\mu_i(k)}\mathbf{A}_3(k))\mathbf{q}_j(k) \nonumber \\
& & -{}^t\mathbf{p}_i(-k)(\mathbf{A}_2(k)+e^{-i\mu_j(k)}\mathbf{A}_0(k))\mathbf{q}_j(k) \nonumber \\
& = & d_i\delta_{ij}
\label{eq29}
\end{eqnarray}
The result of relation (\ref{eq28}) has been used in the case
$e^{i\mu_i(k)}\neq e^{i\mu_j(k)}$
and $d_i$ is a factor depending on the eigenvector $i$.
This gives orthogonality relations on the statevectors associated to the eigenvalues.

\subsection{Absorbing boundary conditions}

Figure~\ref{fig04} presents the periodic medium near the exterior boundary.
In this domain the solution is described by relation (\ref{eq00}), yielding,
respectively for the displacement and force components,
\begin{eqnarray}
\mathbf{q}(x_1,x_2) & = &
\int_{-\frac{\pi}{b_2}}^{\frac{\pi}{b_2}}
\mathbf{\hat{q}}(x_1,k,x_2) e^{ikx_2} dk \nonumber \\
\mathbf{f}(x_1,x_2) & = &
\int_{-\frac{\pi}{b_2}}^{\frac{\pi}{b_2}}
\mathbf{\hat{f}}(x_1,k,x_2) e^{ikx_2} dk
\label{eq30}
\end{eqnarray}
with the force components given by relation (\ref{eq26}).
Introducing the state vector $\mathbf{x} = {}^t({}^t\mathbf{q},{}^t\mathbf{f})$
and decomposing this solution into the different waves, we get
\begin{eqnarray}
\mathbf{x}(x_1,x_2) & = & \int_{-\frac{\pi}{b_2}}^{\frac{\pi}{b_2}}
\mathbf{\hat{x}}(x_1,k,x_2) e^{ikx_2} dk \nonumber \\
& \approx & \int_{-\frac{\pi}{b_2}}^{\frac{\pi}{b_2}}
\sum_{j=1}^{j=2n}a_{j}(x_1,k) \mathbf{x}_j(k) e^{ikx_2} dk
\label{eq31}
\end{eqnarray}
The last relation is the approximation obtained by the finite element computation
of wave solutions presented before.
The condition of outgoing waves means that there is no incoming wave,
so the amplitudes $a_j(x_1,k)$ associated with incoming waves must equal zero.
This condition is obtained by
\begin{equation}
\mathbf{y}_l^-(-k).\sum_{j=1}^{j=2n}a_{j}(x_1,k) \mathbf{x}_j(k) = 0\ \ \
for\ 1\le l\le n^-
\label{eq32}
\end{equation}
In this relation $\mathbf{y}_l^-(-k)$ are the vectors associated to the negative going waves,
given by relation (\ref{eq27}).
Using relation (\ref{eq29}), one gets $a_{j}^-(x_1,k)=0$ for $1\le j\le n^-$
for the amplitudes of the negative going waves.
Introducing the matrix $\mathbf{Y}$ with lines given by $\mathbf{y}_l^-$ leads to
\begin{equation}
\mathbf{Y}(-k).\mathbf{\hat{x}}(x_1,k,x_2) = 0
\label{eq33}
\end{equation}
Decomposing now $\mathbf{\hat{x}}$ into its displacement and force components, doing the same
thing for $\mathbf{Y}(-k)=[\mathbf{Q}(-k) \ \ \mathbf{F}(-k)]$ leads to
\begin{equation}
\mathbf{Q}(-k).\mathbf{\hat{q}}(x_1,k,x_2) +
\mathbf{F}(-k).\mathbf{\hat{f}}(x_1,k,x_2) = 0
\label{eq34}
\end{equation}
The relation on the boundary is
\begin{equation}
\mathbf{\hat{f}}(x_1,k,x_2) = -\mathbf{F}^{-1}(-k)\mathbf{Q}(-k)\mathbf{\hat{q}}(x_1,k,x_2)
\end{equation}
and then from relation (\ref{eq30})
\begin{equation}
\mathbf{f}(x_1,x_2) = -\int_{-\frac{\pi}{b_2}}^{\frac{\pi}{b_2}}
\mathbf{F}^{-1}(-k)\mathbf{Q}(-k)\mathbf{\hat{q}}(x_1,k,x_2) e^{ikx_2} dk
\label{eq35}
\end{equation}
From the inverse relation (\ref{eq01}), one also has
\begin{equation}
\mathbf{\hat{q}}(x_1,k,x_2) =
\frac{b_2}{2\pi} \sum_{m_2=-\infty}^{+\infty} e^{-ik(x_2+m_2b_2)}\mathbf{q}(x_1,x_2+m_2b_2)
\label{eq36}
\end{equation}
which leads to
\begin{equation}
\mathbf{f}(x_1,x_2) = -\frac{b_2}{2\pi}\int_{-\frac{\pi}{b_2}}^{\frac{\pi}{b_2}}
\mathbf{F}^{-1}(-k)\mathbf{Q}(-k)\sum_{m_2=-\infty}^{+\infty} e^{-ik(x_2+m_2b_2)}
\mathbf{q}(x_1,x_2+m_2b_2)dk
\label{eq37}
\end{equation}
Introducing the function
\begin{equation}
\mathbf{G}(x_2) = -\frac{b_2}{2\pi}\int_{-\frac{\pi}{b_2}}^{\frac{\pi}{b_2}}
\mathbf{F}^{-1}(-k)\mathbf{Q}(-k)e^{-ikx_2}dk
\label{eq38}
\end{equation}
The final relation is
\begin{equation}
\mathbf{f}(x_1,x_2) = \sum_{m_2=-\infty}^{+\infty} \mathbf{G}(x_2+m_2b_2)
\mathbf{q}(x_1,x_2+m_2b_2)
\label{eq39}
\end{equation}

This is the impedance relation on the boundary obtained with
the assumption that there is no negative going wave.
This relation is the absorbing boundary condition we were looking for.
It can be computed from the wave vectors and
the force components associated with them.
Relation (\ref{eq39}) involves an infinite number of terms on the boundary.
This relation can also be written as
\begin{eqnarray}
\mathbf{f}(x_1,x_2) & = & \left(\sum_{m_2=-\infty}^{+\infty} \mathbf{G}(x_2+m_2b_2)\right)
\mathbf{q}(x_1,x_2) \nonumber \\
& & + \sum_{m_2=-\infty}^{+\infty} \mathbf{G}(x_2+m_2b_2)
\left(\mathbf{q}(x_1,x_2+m_2b_2)-\mathbf{q}(x_1,x_2)\right) \nonumber \\
& = & \left(\sum_{m_2=-\infty}^{+\infty} \mathbf{G}(x_2+m_2b_2)\right)
\mathbf{q}(x_1,x_2) \nonumber \\
& & + \frac{1}{2b_2}\left(\sum_{m_2=-\infty}^{+\infty} m_2b_2\mathbf{G}(x_2+m_2b_2)\right)
(\mathbf{q}(x_1,x_2+b_2)-\mathbf{q}(x_1,x_2-b_2)) \nonumber \\
& & \displaystyle + \sum_{m_2=-\infty}^{+\infty} \mathbf{G}(x_2+m_2b_2)
[\mathbf{q}(x_1,x_2+m_2b_2)-\mathbf{q}(x_1,x_2) \nonumber \\
& & \displaystyle -\frac{1}{2}m_2(\mathbf{q}(x_1,x_2+b_2)-\mathbf{q}(x_1,x_2-b_2))]
\label{eq39b}
\end{eqnarray}
If $\mathbf{q}$ is slowly varying the last term should be small and
for practical purposes we will use the approximate relations at various orders
given by
\begin{eqnarray}
\mathbf{f}(x_1,x_2) & \approx &\mathbf{G}_0\mathbf{q}(x_1,x_2) +
\frac{\mathbf{G}_1}{2b_2} (\mathbf{q}(x_1,x_2+b_2)-\mathbf{q}(x_1,x_2-b_2)) \nonumber \\
&&+\frac{\mathbf{G}_2}{2b_2^2} (\mathbf{q}(x_1,x_2+b_2)+\mathbf{q}(x_1,x_2-b_2)-2\mathbf{q}(x_1,x_2)) + \dots
\label{eq40}
\end{eqnarray}
with
\begin{eqnarray}
\mathbf{G}_0 & = & \sum_{m_2=-\infty}^{+\infty} \mathbf{G}(x_2+m_2b_2)
= -(\mathbf{F}^{-1}\mathbf{Q})(0) \nonumber \\
\mathbf{G}_1 & = & \sum_{m_2=-\infty}^{+\infty} m_2b_2\mathbf{G}(x_2+m_2b_2)
= i(\mathbf{F}^{-1}\mathbf{Q})'(0) \nonumber \\
\mathbf{G}_2 & = & \sum_{m_2=-\infty}^{+\infty} (m_2b_2)^2\mathbf{G}(x_2+m_2b_2)
= (\mathbf{F}^{-1}\mathbf{Q})''(0)
\label{eq41}
\end{eqnarray}
Relation (\ref{eq40}) involves a finite number of nodes around the point
where the relation is written.
It depends on the number of nodes chosen to approximate
the boundary condition.
This number can be 1 for a crude approximation involving
only one node or can be larger.
For a very large number of nodes, the condition tends towards the true absorbing
condition for a half-plane in the periodic media given by (\ref{eq39}).
Up to now everything has been written for periodic media
but it is clear that homogeneous media are also periodic
media and so all that has been said applies also to homogeneous media.
The condition (\ref{eq40}) can be seen as a generalization of the Taylor
approximation boundary condition proposed by \cite{Engquist77}.
But, while the boundary conditions in \cite{Engquist77} were obtained
by approximation of the exact continuous relations for specific problems, 
they are obtained here directly and with general applicability from 
the discretized equations.

\section{Simple examples}

\subsection{Estimation of the accuracy}

\begin{figure}
  \begin{center}
    \epsfig{file=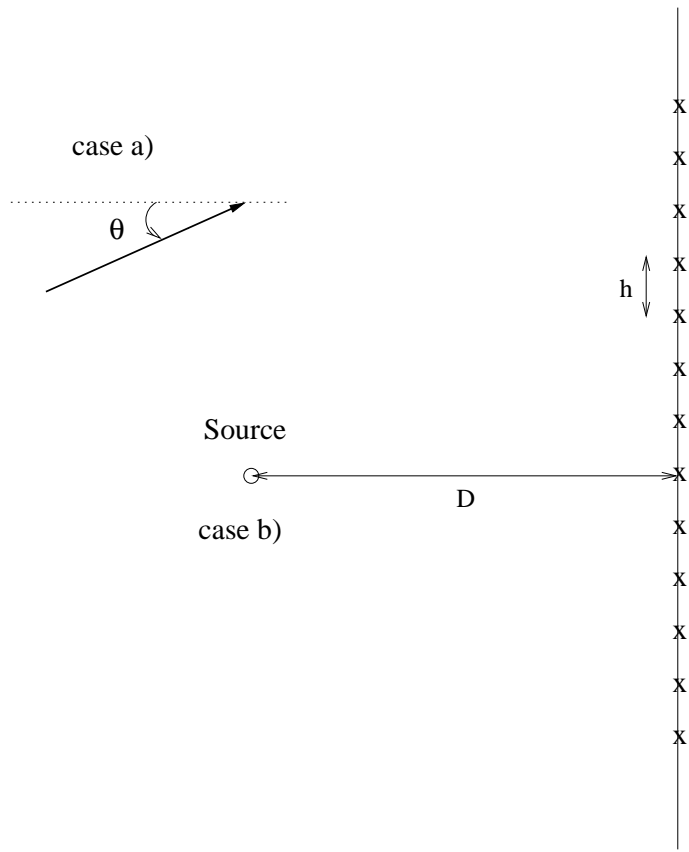,scale=.7}
    \caption{Points used to estimate the boundary condition with sound pressures created
by a plane wave (case a) and a point source (case b)}
    \label{fig05}
  \end{center}
\end{figure}

In this section we try to estimate the quality of the proposed
boundary condition compared with known relations for the simple case
of the two-dimensional acoustics.
Consider first a plane wave
incident on the plane $y=0$ at an angle $\theta$
with the normal to the plane.
Let us define points at a horizontal distance $D$ from the origin
and with a vertical spacing $h$, see figure \ref{fig05}.
The sound pressure at point $(D,lh)$ is given by
\begin{equation}
p_a^l = e^{iK(\cos \theta D + \sin \theta lh)}
\label{eq65}
\end{equation}
where $K = \omega/c$ is the wavenumber and $c$ is the sound velocity.
The analytical force at the same points is the normal derivative
in direction 1 given by
\begin{equation}
f^l_a = iK\cos \theta e^{iK(\cos \theta D + \sin \theta lh)}
\label{eq66}
\end{equation}

For a point source at origin, the pressure is solution of
\begin{equation}
  \Delta p + K^2p = -\delta(r)
\label{eq43}
\end{equation}
where $r$ is the distance from the origin
and $\delta(.)$ is the Dirac function. The solution of this
equation for the time dependence $e^{-i\omega t}$ is given by
\begin{equation}
  G(r) = \frac{i}{4}H_0(Kr)
\label{eq44}
\end{equation}
where $H_0$ is the Hankel function of zero order and first type.
The analytical solution at each point $l$ is
\begin{equation}
 p^l_a = \frac{i}{4}H_0(K\sqrt{D^2+(lh)^2})
\label{eq45}
\end{equation}
and the analytical force at the same points is the normal derivative
(in direction 1) given by
\begin{equation}
 f^l_a = -\frac{iKD}{4\sqrt{D^2+(lh)^2}}H_1(K\sqrt{D^2+(lh)^2})
\label{eq46}
\end{equation}
The absorbing boundary condition described in the precedent section
will allow to compute numerical forces $f^{l}_n$ at a node
from the knowledge of $p^l_a$. If the boundary condition was perfect
one would have $f^{l}_n=f^{l}_a$ but the proposed condition is approximate
and one only has $f^{l}_n \approx f^{l}_a$.
The error can be estimated by
\begin{equation}
 e = \frac{\displaystyle |f^{l}_n-f^{l}_a|}
{\displaystyle |f^{l}_a|}
\label{eq47}
\end{equation}
The next step is to compute $f^l_n$ from the method
proposed in this paper and the error by relation (\ref{eq47}) to estimate
the quality of the absorbing condition.

\subsection{Acoustic element}

Consider the rectangular four nodes acoustic element of size $b_1\times b_2$.
The elementary stiffness and mass matrices are given by
\begin{equation}
  \mathbf{K} =
  \frac{1}{6b_1b_2}
  \left [
    \begin{array}{cccc}
      (2b_2^2+2b_1^2) & (-2b_2^2+b_1^2) & (-b_2^2-b_1^2) & (b_2^2-2b_1^2)\\
      (-2b_2^2+b_1^2) & (2b_2^2+2b_1^2) & (b_2^2-2b_1^2) & (-b_2^2-b_1^2)\\
      (-b_2^2-b_1^2) & (b_2^2-2b_1^2) & (2b_2^2+2b_1^2) & (-2b_2^2+b_1^2)\\
      (b_2^2-2b_1^2) & (-b_2^2-b_1^2) & (-2b_2^2+b_1^2) & (2b_2^2+2b_1^2)
    \end{array}
  \right ]
\label{eq48}
\end{equation}
\begin{equation}
  \mathbf{M} =
  \frac{b_1b_2}{36c^2}
  \left [
    \begin{array}{cccc}
      4 & 2 & 1 & 2\\
      2 & 4 & 2 & 1\\
      1 & 2 & 4 & 2\\
      2 & 1 & 2 & 4
    \end{array}
  \right ]
\label{eq49}
\end{equation}

and the dynamic stiffness matrix can then be determined by
$\mathbf{D} = \mathbf{K}-\omega^2\mathbf{M}$.

It can be noted that the reduced set of displacement dofs $\mathbf{q}_r$
contains only $\mathbf{q}_{LB}$. Then, the matrices $\mathbf{W}_0$ and
$\mathbf{W}_1$ have the following forms:
\begin{equation}
 \mathbf{W}_0 = {}^t\left[1 \quad 0 \quad 0 \quad e^{ikb_2}\right]\qquad;\qquad
 \mathbf{W}_1 = {}^t\left[0 \quad 1 \quad e^{ikb_2} \quad 0\right]
\label{eq50}
\end{equation}

The terms $\mathbf{A}_j$ in equation (\ref{eq21}) are given by
\begin{eqnarray}
  A_0(k) &=& \mathbf{W}_1^*(k)\mathbf{D}_l(k)\mathbf{W}_0(k) \nonumber \\
  &=& -\frac{1}{18}\left [12\frac{b_2}{b_1}-6\frac{b_1}{b_2}+2K^2b_1b_2+\left
      (6\frac{b_2}{b_1}+6\frac{b_1}{b_2}+K^2b_1b_2\right )\cos(kb_2)\right ]\nonumber \\
  A_1(k) &=& \mathbf{W}_0^*(k)\mathbf{D}_l(k)\mathbf{W}_0(k) \nonumber \\
  &=& -\frac{1}{9}\left [-6\frac{b_2}{b_1}-6\frac{b_1}{b_2}+2K^2b_1b_2+\left
      (-3\frac{b_2}{b_1}+6\frac{b_1}{b_2}+K^2b_1b_2\right )\cos(kb_2)\right ]\nonumber \\
  A_2(k) &=& \mathbf{W}_1^*(k)\mathbf{D}_l(k)\mathbf{W}_1(k) \nonumber \\
  &=& -\frac{1}{9}\left [-6\frac{b_2}{b_1}-6\frac{b_1}{b_2}+2K^2b_1b_2+\left
      (-3\frac{b_2}{b_1}+6\frac{b_1}{b_2}+K^2b_1b_2\right )\cos(kb_2)\right ]\nonumber \\
  &=& A_1(k) \nonumber \\
  A_3(k) &=& \mathbf{W}_0^*(k)\mathbf{D}_l(k)\mathbf{W}_1(k)\nonumber\\
  &=& -\frac{1}{18}\left [12\frac{b_2}{b_1}-6\frac{b_1}{b_2}+2K^2b_1b_2+\left
      (6\frac{b_2}{b_1}+6\frac{b_1}{b_2}+K^2b_1b_2\right
    )\cos(kb_2)\right ]\nonumber\\
  &=& A_0(k)
\label{eq51}
\end{eqnarray}

The eigensolutions of the spectral problem (\ref{eq20}) are
then determined and are given by
\begin{eqnarray}
  e^{i\mu} &=& \frac{1}{2A_3}\left (-(A_1+A_2)\pm \sqrt{(A_1+A_2)^2-4A_0^2}\right
  )\label{eq52}\\
  e^{-i\mu} &=& \frac{1}{2A_3}\left (-(A_1+A_2)\mp \sqrt{(A_1+A_2)^2-4A_0^2}\right
  )\label{eq53}
\end{eqnarray}
The signs are selected as in section 2.3. As there is only one dof
in this case, one has $n^+=1$ and after normalization one can choose
$\mathbf{q}_1(k) = 1$. From relation (\ref{eq27}), taking also
$\mathbf{p}_j(-k)=1$, yields
\begin{eqnarray}
y(-k) & = & \left[
\begin{array}{cc}
A_2+e^{i\mu}A_3 \ \ & \ \ 1
\end{array} \right] \nonumber \\
& = & \left[
\begin{array}{cc}
\frac{A_3}{2A_3}\left (-(A_1+A_2)\pm \sqrt{(A_1+A_2)^2-4A_0^2}\right)
+A_2 \ \ & \ \ 1
\end{array} \right] \nonumber \\
& = & \left[
\begin{array}{cc}
\pm \frac{1}{2}\sqrt{(A_1+A_2)^2-4A_0^2} \ \ & \ \ 1  \nonumber \\
\end{array} \right]
\label{eq54}
\end{eqnarray}
this gives, with the notation of relation (\ref{eq34}),
\begin{eqnarray}
Q(-k) & = & \pm \frac{1}{2}\sqrt{(A_1+A_2)^2-4A_0^2} \nonumber \\
F(-k) & = & 1
\label{eq55}
\end{eqnarray}

Near $k=0$ (this means in fact near the normal incidence)
one has the development
\begin{equation}
  \left\{
    \begin{array}{rcl}
      A_0(k) &=& \displaystyle-\frac{b_2}{b_1}-\frac{1}{6}K^2b_1b_2
+\frac{1}{36}\left(6\frac{b_2}{b_1}+6\frac{b_1}{b_2}+K^2b_1b_2\right)(kb_2)^2+O((kb_2)^4)\\
      A_1(k) &=& \displaystyle \frac{b_2}{b_1}-\frac{1}{3}K^2b_1b_2
+\frac{1}{18}\left(-3\frac{b_2}{b_1}+6\frac{b_1}{b_2}+K^2b_1b_2\right)(kb_2)^2+O((kb_2)^4)\\
      A_2(k) &=& A_1(k)\\
      A_3(k) &=& A_0(k)
    \end{array}
  \right.
\label{eq56}
\end{equation}
and this leads to
\begin{eqnarray}
Q(-k) & = & \pm iKb_2\sqrt{1-\frac{1}{12}(Kb_1)^2} \times \nonumber \\
& & \displaystyle \left(1 -\frac{1}{2}\frac{1+\frac{(Kb_2)^2}{3}-\frac{(Kb_1)^2}{6}-\frac{(K^2b_1b_2)^2}{36}}{1-\frac{(Kb_1)^2}{12}}\frac{k^2}{K^2}\right)+O((kb_2)^4)
\label{eq57}
\end{eqnarray}

From relation (\ref{eq26}) one also has
\begin{eqnarray}
  f^r(k) & = & \mathbf{W}_0^*\mathbf{D}_l\left (\mathbf{W}_0 +
    e^{i\mu}\mathbf{W}_1\right ) \nonumber \\
  & = & A_1 + e^{i\mu}A_3 \nonumber \\
  & = & Q(-k)
  \label{eq58}
\end{eqnarray}
Following the rule that the positive waves are such that
$\mathrm{Re}\left\{i\omega\mathbf{q}^H_j\mathbf{f}_j^r\right\}>0$,
one has to choose the minus sign in relation (\ref{eq55}).
For the case $Kb_1\ll 1$ and $k\ll K$, one has the approximation:
\begin{equation}
  f(0) \simeq -iKb_2
  \label{eq59}
\end{equation}
The power across the boundary is thus
\begin{equation}
P = \frac{1}{2} Re(f_{left \rightarrow right}v^*)
= \frac{1}{2} Re(f^r(0)(-i\omega)^*)
= \frac{1}{2} Kb_2\omega > 0
\label{eq60}
\end{equation}

For the second order approximation one has
\begin{eqnarray}
G_0 & = & -(F^{-1}Q)(0)
= iKb_2\sqrt{1-\frac{1}{12}(Kb_1)^2} \approx iKb_2 \nonumber \\
G_1 & = & i(F^{-1}Q)'(0) = 0 \nonumber \\
G_2 & = & (F^{-1}Q)''(0)
= \frac{ib_2}{K}\frac{1+\frac{(Kb_2)^2}{3}-\frac{(Kb_1)^2}{6}-\frac{(K^2b_1b_2)^2}{36}}{\sqrt{1-\frac{1}{12}(Kb_1)^2}} \approx \frac{ib_2}{K}
\label{eq62}
\end{eqnarray}

The relation between forces and displacements dofs on the boundary
of the element is thus given by using (\ref{eq40})
\begin{eqnarray}
f(x_1,x_2) &\approx &iKb_2q(x_1,x_2) \nonumber \\
&& +\frac{i}{2Kb_2} (q(x_1,x_2+b_2)+q(x_1,x_2-b_2)-2q(x_1,x_2))
\label{eq63}
\end{eqnarray}
At order 0 one finds the classical approximation
of the radiating boundary condition.
The factor $b_2$ is present because the force is calculated over an edge
of an element of length $b_2$.

We compare four solutions in the following
\begin{enumerate}
\item The zero order solution with the numerical computation of $Q(0)$
leading to the relation $\displaystyle f^l_n = G_0p_a^l $
\item The zero order solution with the simplified computation of $Q(0)$
leading to the relation $\displaystyle f^l_n = iKb_2p_a^l $
\item The second order solution with the numerical computation of
$G_2=(F^{-1}Q)''(0) \approx ((F^{-1}Q)(\delta) + (F^{-1}Q)(-\delta)-2(F^{-1}Q)(0))/\delta^2$
leading to the relation
$\displaystyle f^l_n = G_0p_a^l +\frac{G_2}{2b_2^2} (p_a^{l+1}+p_a^{l-1}-2p_a^l)$
\item The second order solution with the simplify computation of $Q(0)$
and $(F^{-1}Q)''(0)$ leading to the relation
$f^l_n = iKb_2p_a^l+\frac{i}{2Kb_2} (p_a^{l+1}+p_a^{l-1}-2p_a^l)$
\end{enumerate}

\subsection{Example}

Here we compute the error of relation (\ref{eq47})
for different cases as shown in figure \ref{fig05}.
Case a) is for a sound pressure created by a plane wave while in case b)
the sound pressure is created by a point source.
The acoustic element used for the computation of the boundary condition
can be of size $b_1\times b_2=0.01 m \times 0.01 m$
or $b_1\times b_2=0.05 m \times 0.05 m$.
The sound velocity is $c=340 m/s$ and the distance between the points is $h=b_2$.

\begin{figure}
  \begin{center}
(a) \\
    \epsfig{file=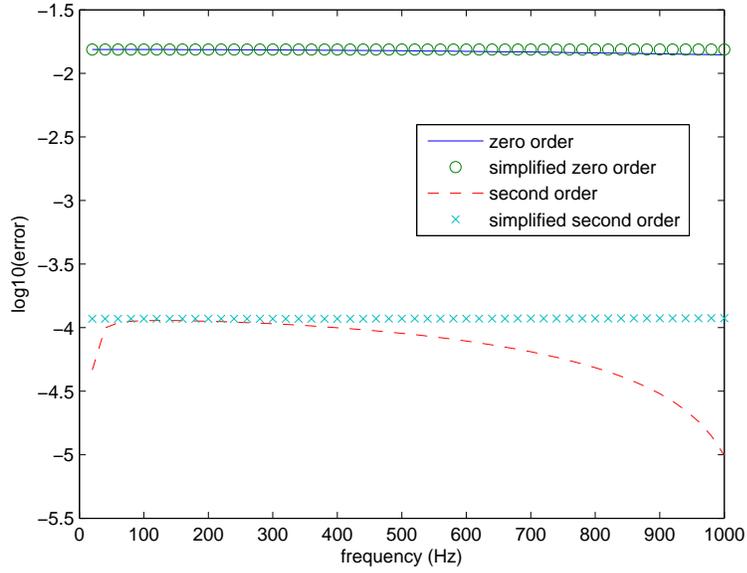,scale=.7} \\
(b) \\
    \epsfig{file=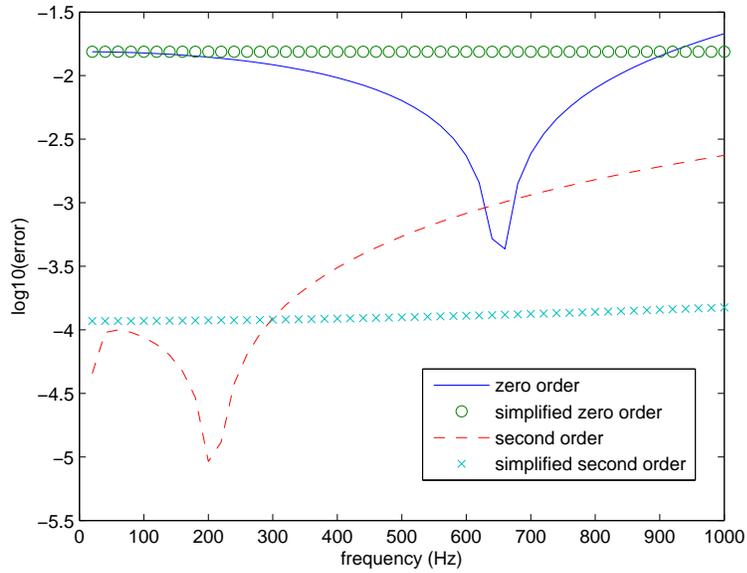,scale=.7}
    \caption{Error versus the frequency for a plane wave at $10^o$
for (a) an element size $0.01m \times 0.01m$ and (b) an element size $0.05m \times 0.05m$}
    \label{fig06}
  \end{center}
\end{figure}

The first example is for a pressure created by a plane wave at the incidence angle $\theta = 10^o$.
The error for the four cases listed in the precedent section are plotted in figure \ref{fig06}.
The error is the same for any point on the vertical axis.
It can be observed that the second order relations are much better than the first ones as expected.
The comparison of the two sizes for the acoustic element shows that the size
$0.05m\times 0.05m$ can reduce the accuracy of the solution for high frequencies
and the second order condition.
In these cases it is better to use elements with small sizes.

\begin{figure}
  \begin{center}
    \epsfig{file=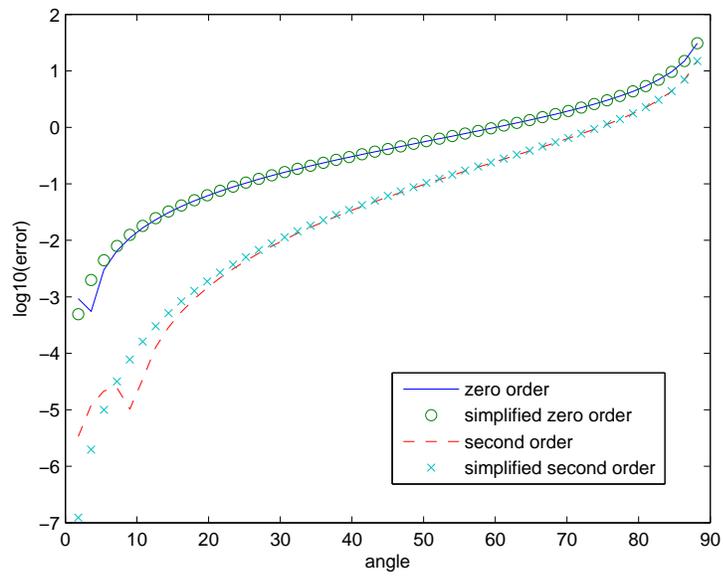,scale=.7}
    \caption{Error versus the angle of incidence}
    \label{fig07}
  \end{center}
\end{figure}

In figure \ref{fig07} the error is plotted versus the angle of incidence of the plane wave.
An acoustic element of size $b_1\times b_2=0.01 m \times 0.01 m$ has been used.
The solution is accurate (error less than 1\%) for angles up to $10^o$ for a zero order
condition and up to $30^o$ for a second order condition.
This clearly shows that second order conditions are much better for waves
at oblique incidence.

\begin{figure}
  \begin{center}
    \epsfig{file=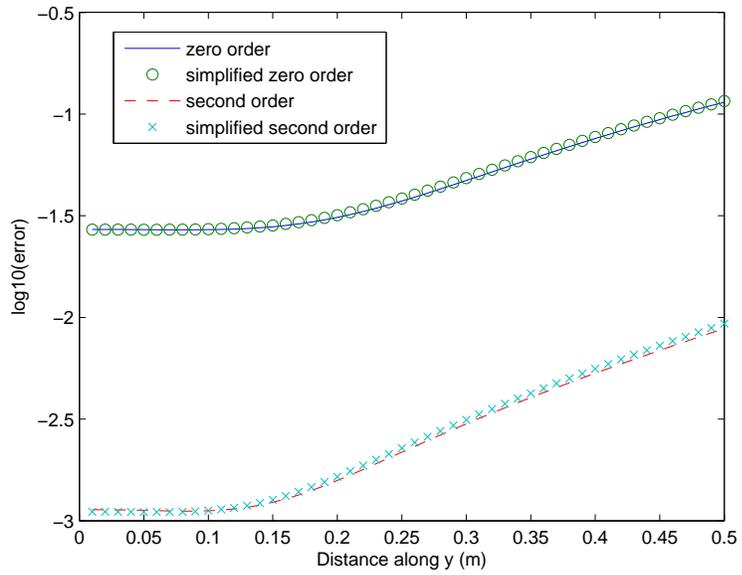,scale=.7}
    \caption{Error versus the distance for a point source}
    \label{fig08}
  \end{center}
\end{figure}

Finally the error is plotted versus the distance along the $y$ axis in figure \ref{fig08}
for a pressure created by a point source at distance $D=1m$ on the $x$ axis and
at the frequency $1000Hz$. The reduction in accuracy can be observed as we
move along the $y$ axis leading to greater incidence angles in agreement
with the precedent observation on the plane waves.
All these points confirm the accuracy of the method proposed here.


\section{Finite element examples}

\subsection{Acoustics}

\begin{figure}
  \begin{center}
    \epsfig{file=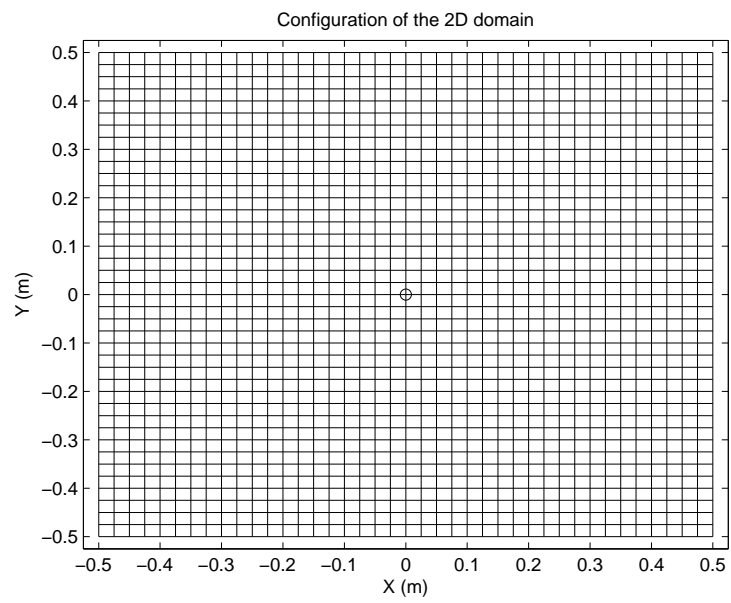,scale=.7}
    \caption{Example of finite element domain with a point source in its center.}
    \label{fig09}
  \end{center}
\end{figure}

In this section we use the precedent boundary condition to solve
some finite element problems for different frequencies and mesh densities.
We consider first a finite element acoustic problem on a square domain
with a point source excitation in its center.
A two dimensional domain of size $1m\times 1m$ is generated by Ansys.
The domain and an example of mesh are presented in figure \ref{fig09}.
The size of the acoustic element is $b = 0.025m$
leading to $40 \times 40$ elements for the whole domain.
In each case, only square elements are used for the mesh.
The sound velocity is $c=340m/s$.
Only mesh information, stiffness and mass matrices are picked
out and are then introduced into Matlab to get the results presented below.
The procedure is done over the frequency band $[0,2000Hz]$.

\begin{figure}
  \begin{center}
a) \\
    \epsfig{file=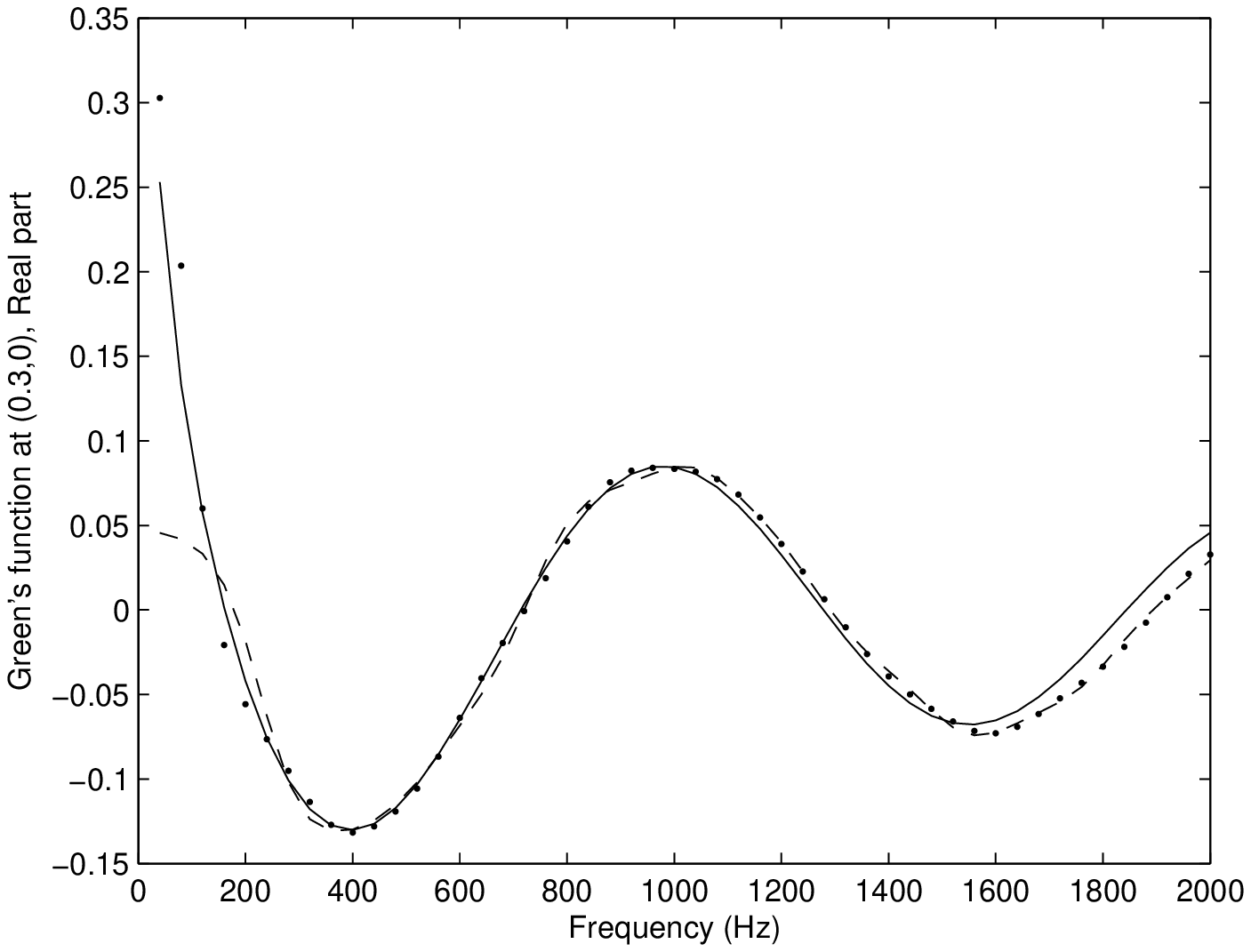,scale=.7} \\
b) \\
    \epsfig{file=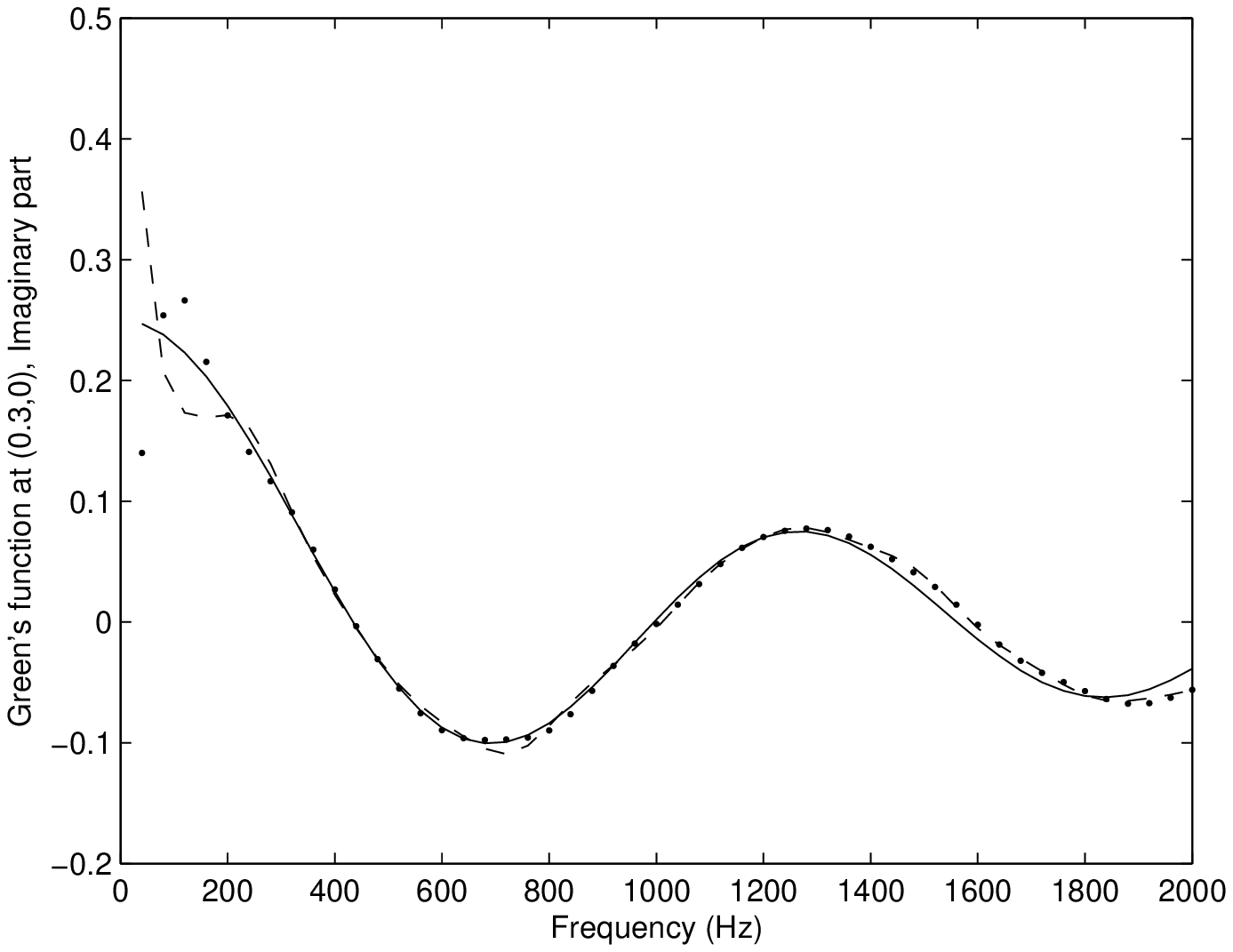,scale=.7}
    \caption{Comparison of analytical $\line(1,0){15}$ and numerical Green's functions,
      with the present method at order 0 $--$ and at order 2 $.\ .$ for the 2D acoustics
      at point $(0.3,0)$ with $L=1m$ and $b=0.025m$:
      a) real part, b) imaginary part.\label{fig10}}
  \end{center}
\end{figure}

\begin{figure}
  \begin{center}
a) \\
    \epsfig{file=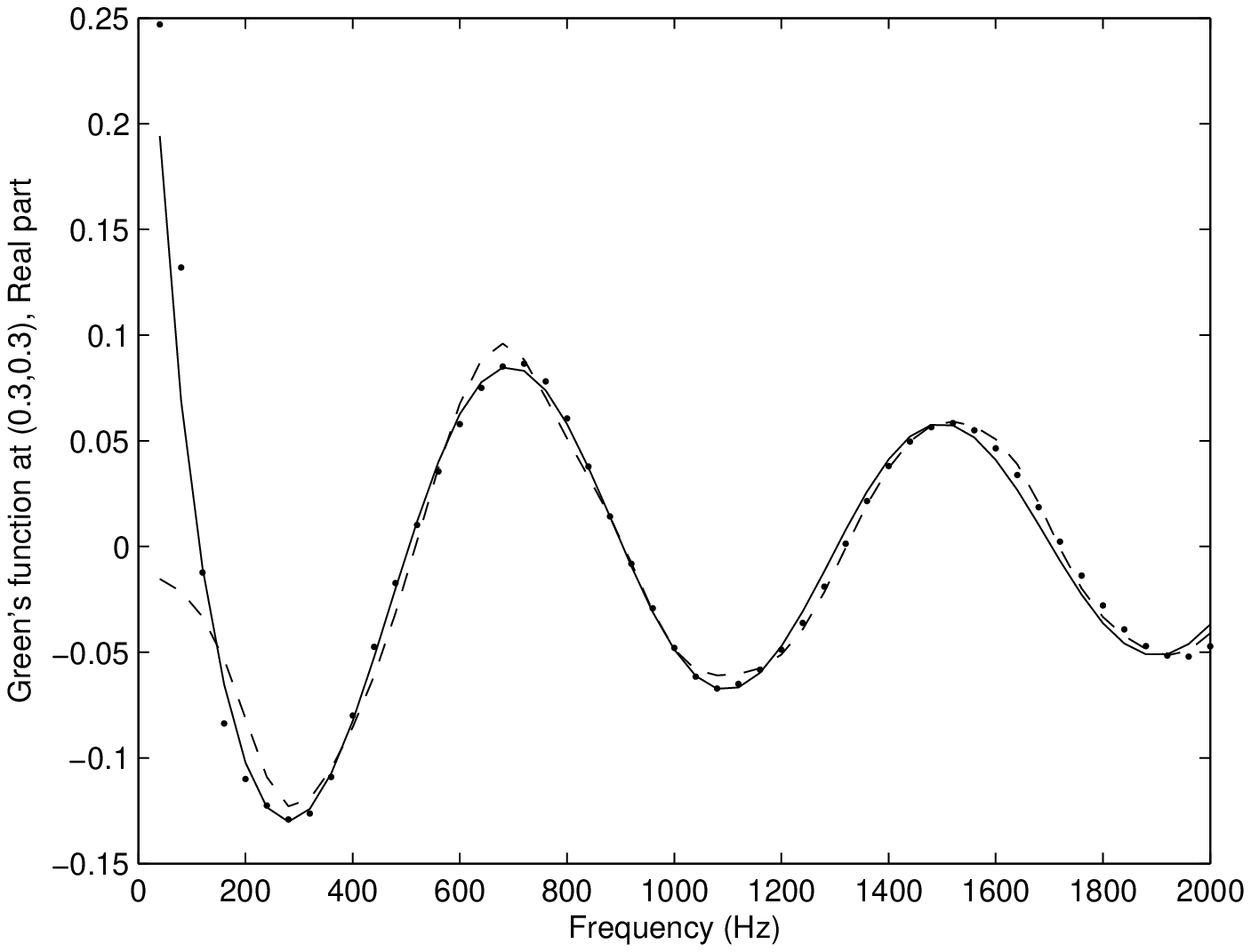,scale=.7} \\
b) \\
    \epsfig{file=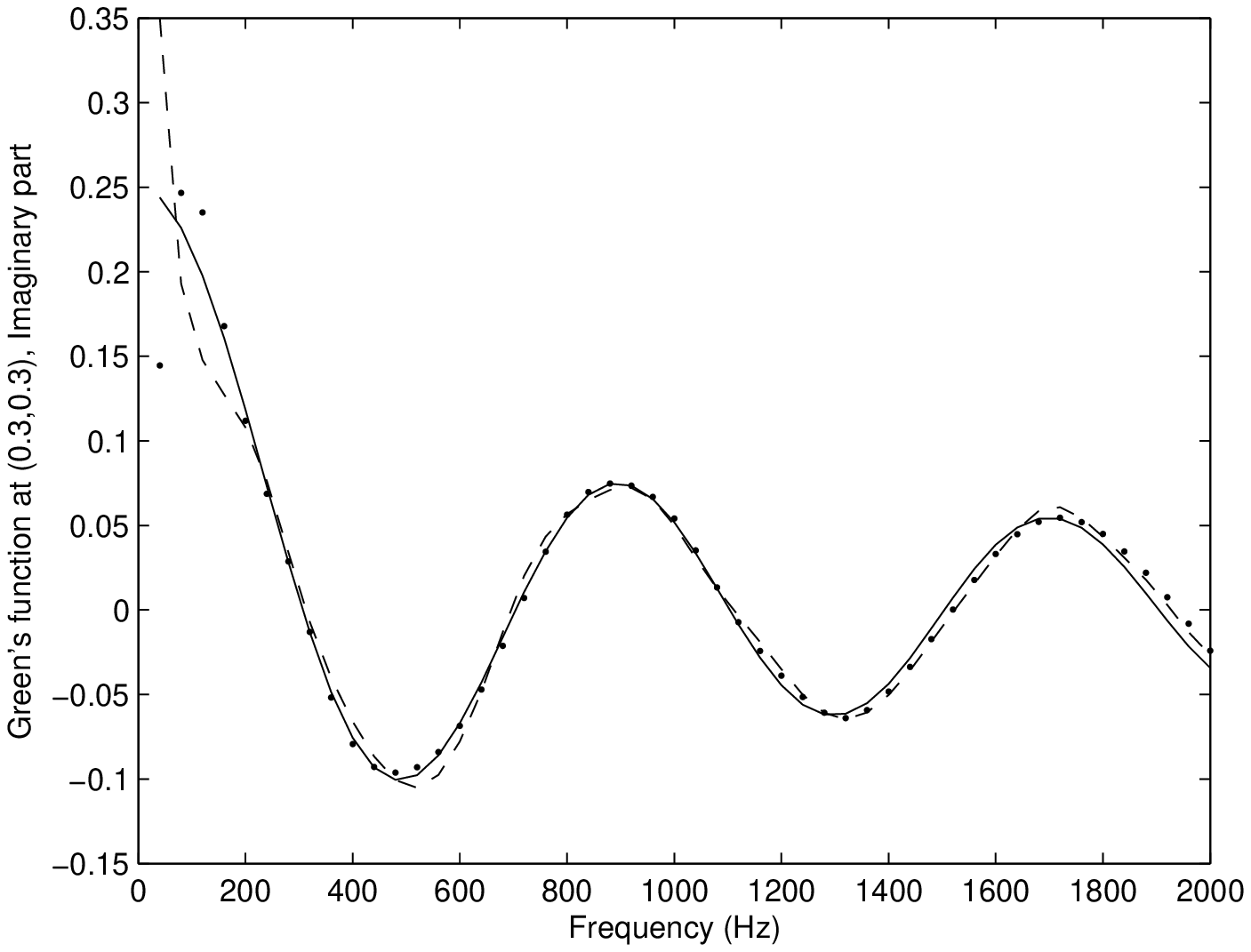,scale=.7}
    \caption{Comparison of analytical $\line(1,0){15}$ and numerical Green's functions,
      with the present method at order 0 $--$ and at order 2 $.\ .$ for the 2D acoustics
      at point $(0.3,0.3)$ with $L=1m$ and $b=0.025m$:
      a) real part, b) imaginary part.\label{fig11}}
  \end{center}
\end{figure}

\begin{figure}
  \begin{center}
    \epsfig{file=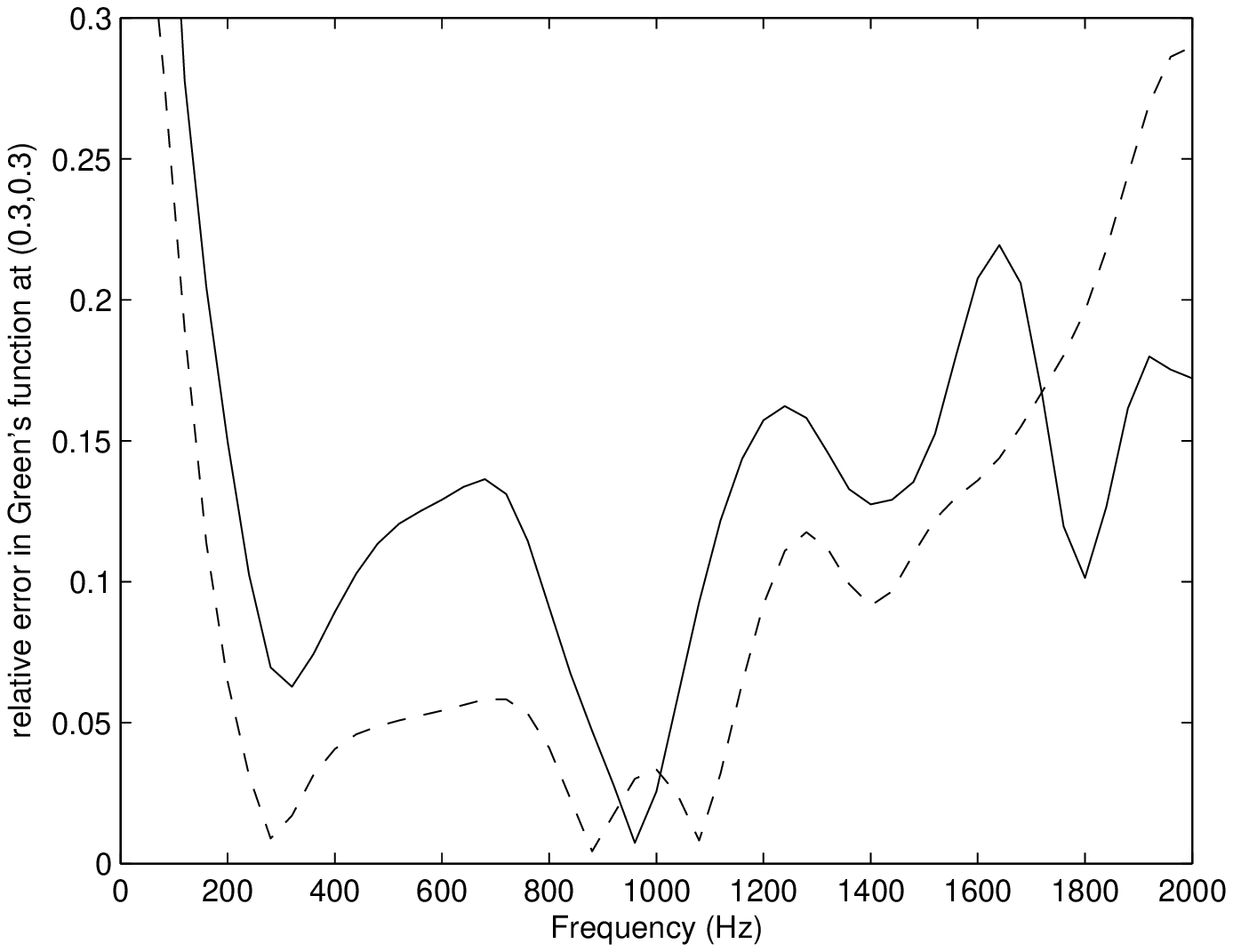,scale=.7}
    \caption{Comparison of relative errors for 0 order $\line(1,0){15}$ and
      second order $--$ boundary conditions for the 2D acoustics
      at point $(0.3,0.3)$ with $L=1m$ and $b=0.025m$.\label{fig12}}
  \end{center}
\end{figure}

Numerical Green's functions are calculated for zero and second order boundary conditions.
The excitation is at point $(0,0)$ and the analytical solution
in infinite space is given by formula (\ref{eq45}).
The Green's functions are presented in figure \ref{fig10}
for a point at $(0.3,0)$ on the horizontal axis
and in figure \ref{fig11} for a point at $(0.3,0.3)$ along the diagonal.
Good agreements between the two types of absorbing boundary conditions
and the analytical solution can be observed.
Both boundary conditions fail at low frequencies because
the size of the domain is too small compared to the wavelength.
Similarly the error for high frequencies are of same orders for both boundary conditions.
For intermediate frequencies the error is lower for the second order boundary condition.
This is more clearly seen in figure \ref{fig12} where the relative error
for the point $(0.3,0.3)$ is plotted versus the frequency.

\begin{figure}
  \begin{center}
a) \\
    \epsfig{file=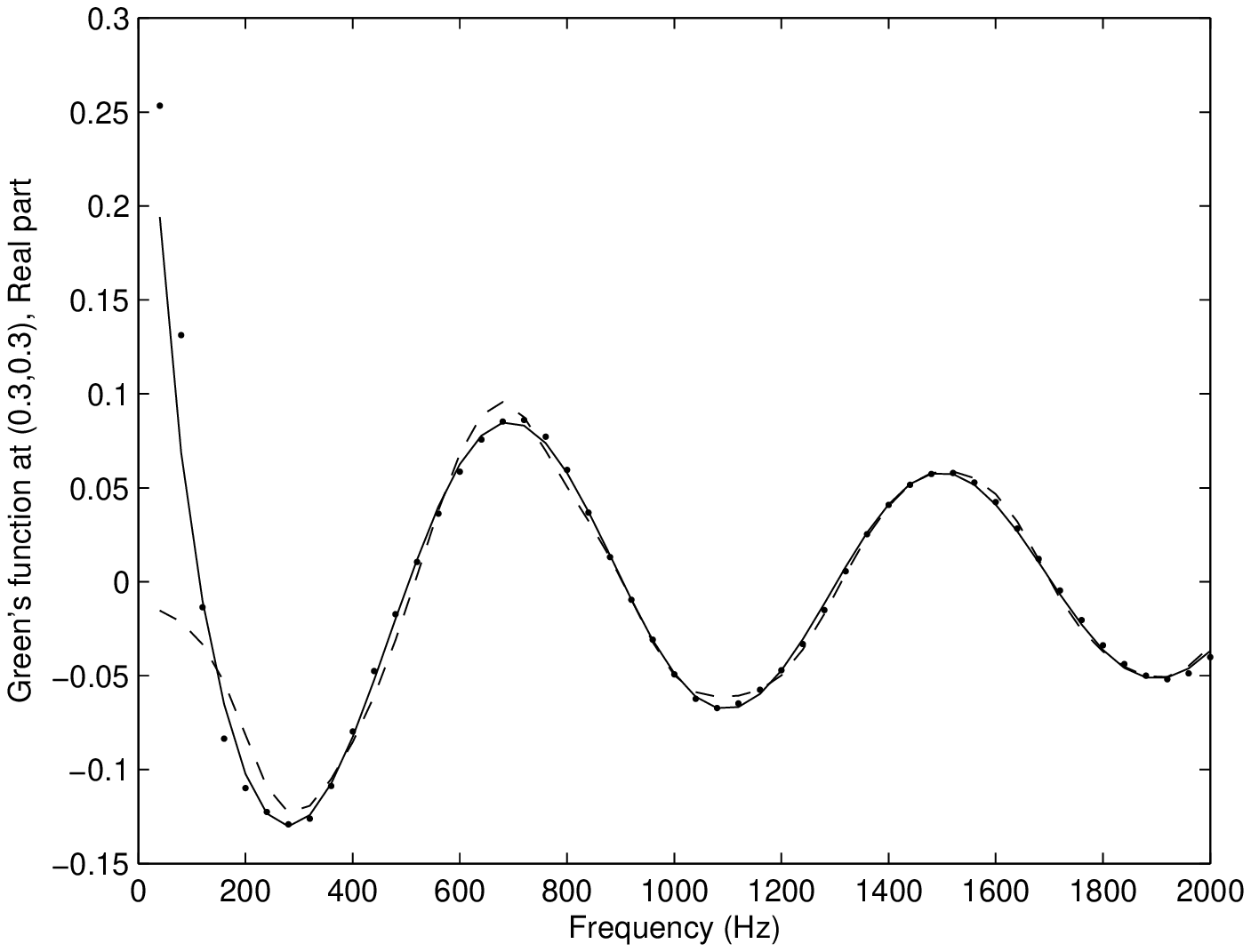,scale=.7} \\
b) \\
    \epsfig{file=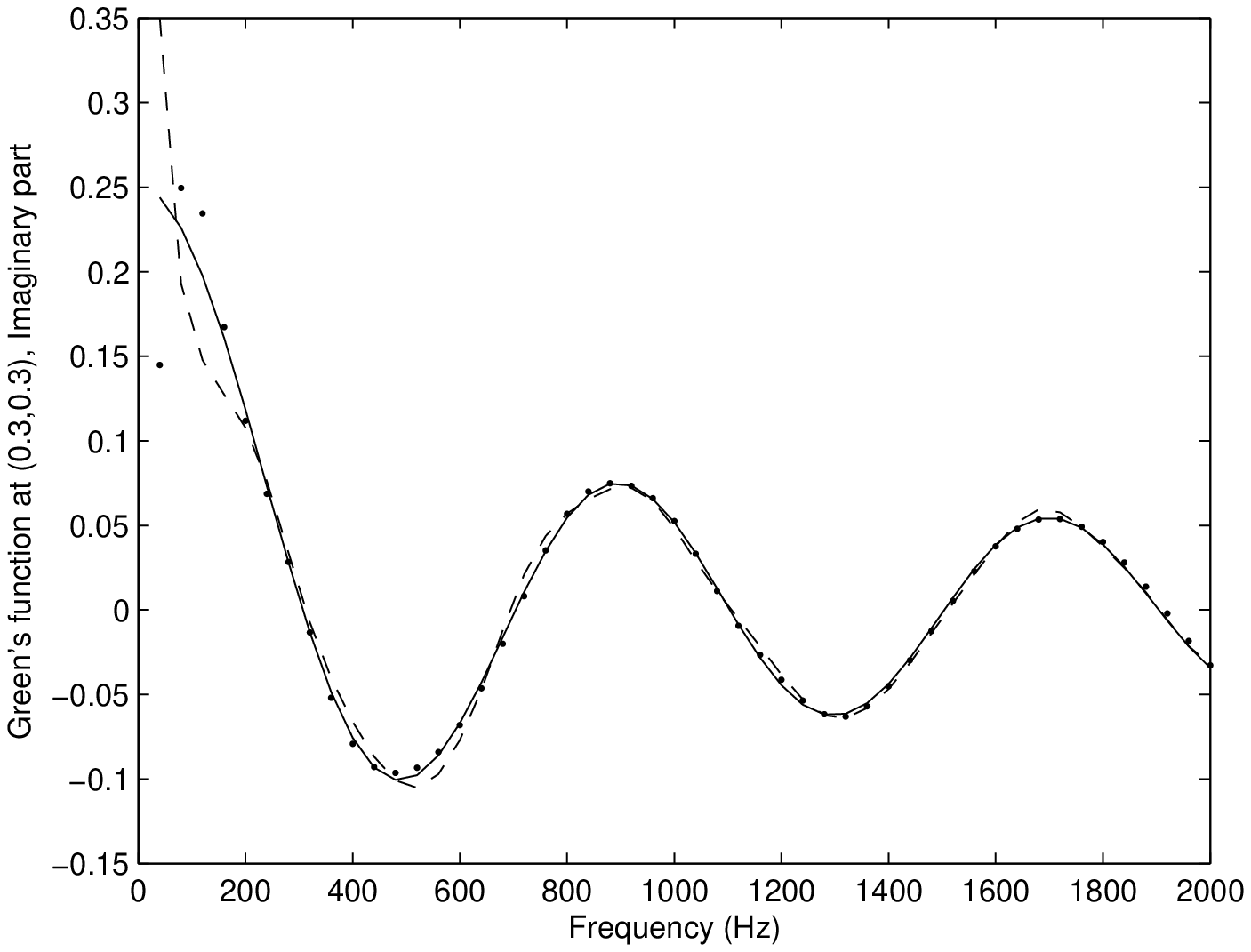,scale=.7}
    \caption{Comparison of analytical $\line(1,0){15}$ and numerical Green's functions,
       with the present method at order 0 $--$ and at order 2 $.\ .$ for the 2D acoustics
      at point $(0.3,0.3)$ with $L=1m$ and $b=0.0125m$:
      a) real part, b) imaginary part.\label{fig13}}
  \end{center}
\end{figure}

\begin{figure}
  \begin{center}
a) \\
    \epsfig{file=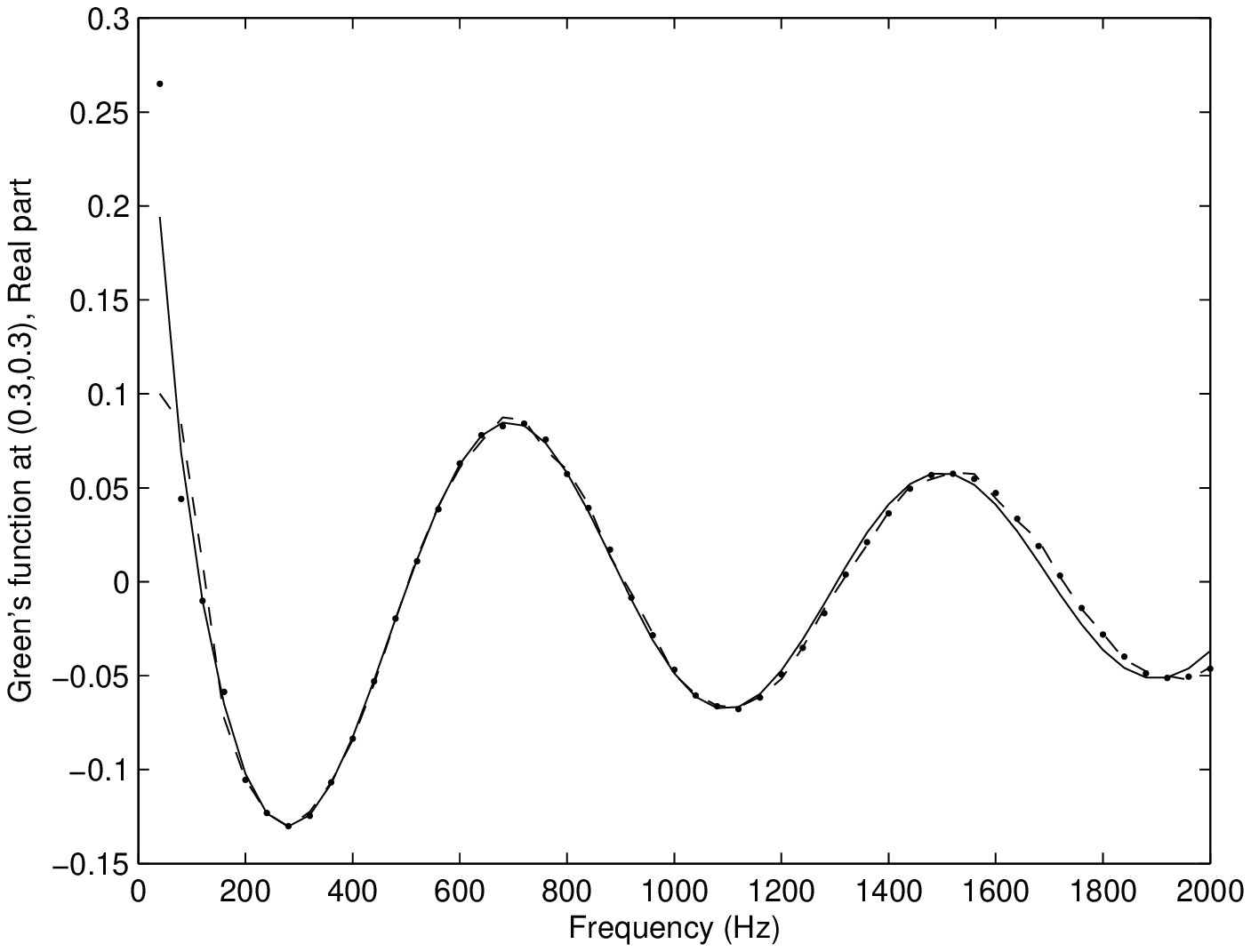,scale=.7} \\
b) \\
    \epsfig{file=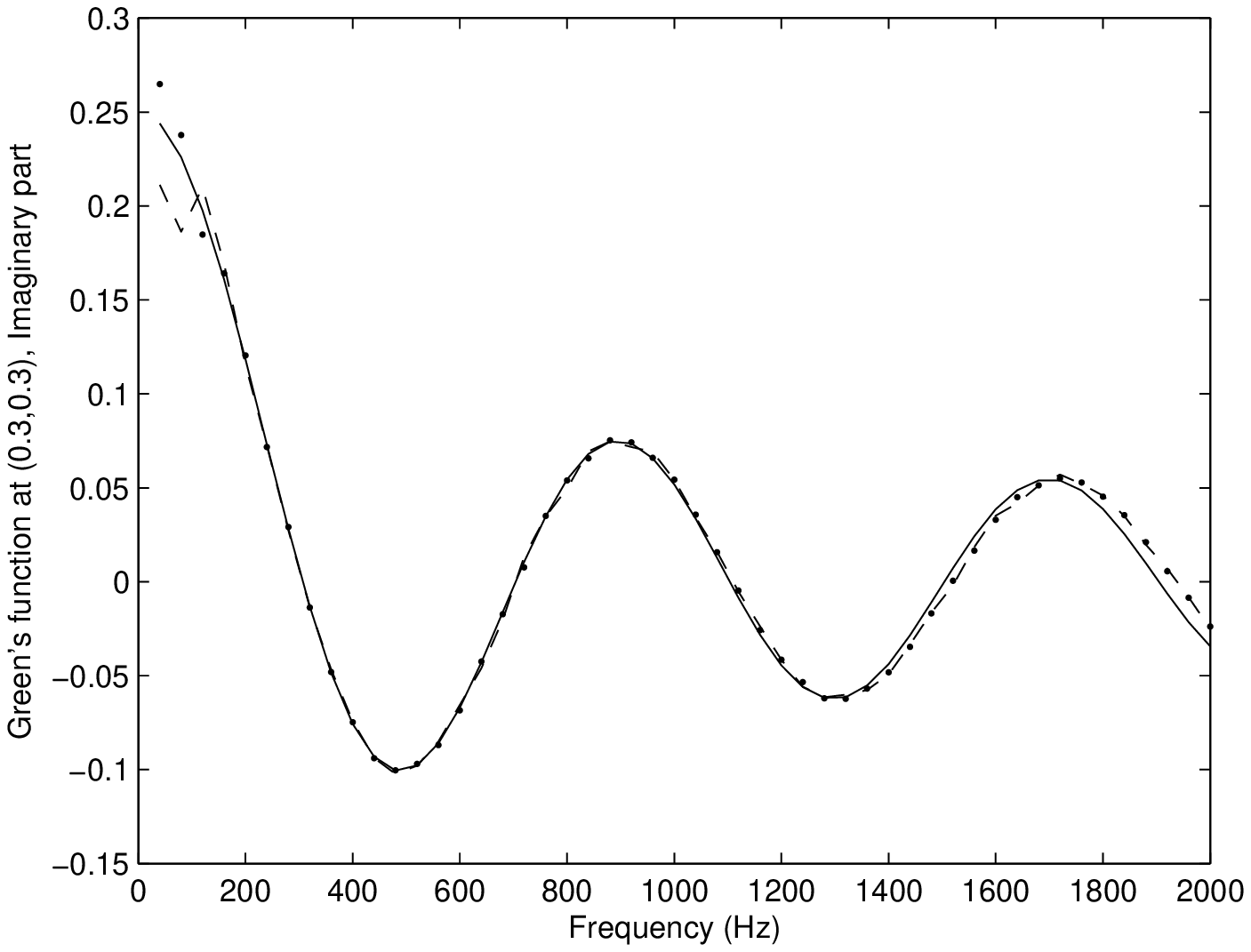,scale=.7}
    \caption{Comparison of analytical $\line(1,0){15}$ and numerical Green's functions,
      with the present method at order 0 $--$ and at order 2 $.\ .$ for the 2D acoustics
      at point $(0.3,0.3)$ with $L=2m$ and $b=0.025m$:
      a) real part, b) imaginary part.\label{fig14}}
  \end{center}
\end{figure}

The same results are presented in figure \ref{fig13} for the point $(0.3,0.3)$
and a mesh density of $80 \times 80$ elements.
The solution is clearly much better at high frequencies meaning
that the errors seen in figures \ref{fig10} and \ref{fig11} can be explained
by elements too large for these frequencies and not by the quality
of the boundary conditions.
In figure \ref{fig14} the domain is now $2m \times 2m$ with $80 \times 80$ elements,
so with elements of the same size as for figures \ref{fig10} and \ref{fig11}.
Results are plotted for the point $(0.3,0.3)$.
Now the improvement is clearly seen for low frequencies
while there is no difference for high frequencies.

\subsection{Two dimensional elastodynamics}

The analytical solution in direction $e_n$ of the two dimensional elastodynamics case when
submitted to a unit force at origin in direction $e_m$, is given by
\begin{equation}
  G_{nm}(r) = \frac{i}{4\mu}\left[A\delta_{nm} + B\frac{x_nx_m}{r^2}\right]
\end{equation}
with:
\begin{eqnarray*}
  A &=& H_0(K_Tr)-\frac{1}{K_Tr}\left[H_1(K_Tr)-\beta
    H_1(K_Lr)\right]\\
  B &=& -2A + \left[H_0(K_Tr) + \beta^2H_0(K_Lr)\right]
\end{eqnarray*}
where $H_0$ and $H_1$ are the Hankel functions of first type, of orders
zero and one respectively.
The wavenumbers are $K_L=\omega/c_L$ and $K_T=\omega/c_T$ for
the longitudinal and transverse waves respectively.
The velocities $c_L$, $c_T$ and the ratio between them $\beta$ are given by,
\begin{equation}
  \left \{
    \begin{array}{rcl}
      \beta &=& \displaystyle\frac{c_T}{c_L}\\
      c_L^2 &=& \displaystyle\frac{\lambda+2\mu}{\rho}\\
      c_T^2 &=& \displaystyle\frac{\mu}{\rho}
    \end{array}
  \right.
  \quad \mathrm{with} \quad
  \lambda=\frac{E\nu}{(1+\nu)(1-2\nu)} \quad \mathrm{and}\quad
  \mu=\frac{E}{2(1+\nu)}
\end{equation}

In this example, the same global meshes as for the acoustic case are
used. The boundary condition is computed from the square four nodes
elements. The material is steel with $E=2.10^{11}Pa$, $\nu = 0.3$
and $\rho = 7800kg/m^3$ and plane strain conditions are used in the
computation. The sizes of the periodic cell can be $b =0.025m,
\,0.0125m$ or $0.00625m$.

\begin{figure}
  \begin{center}
    \epsfig{file=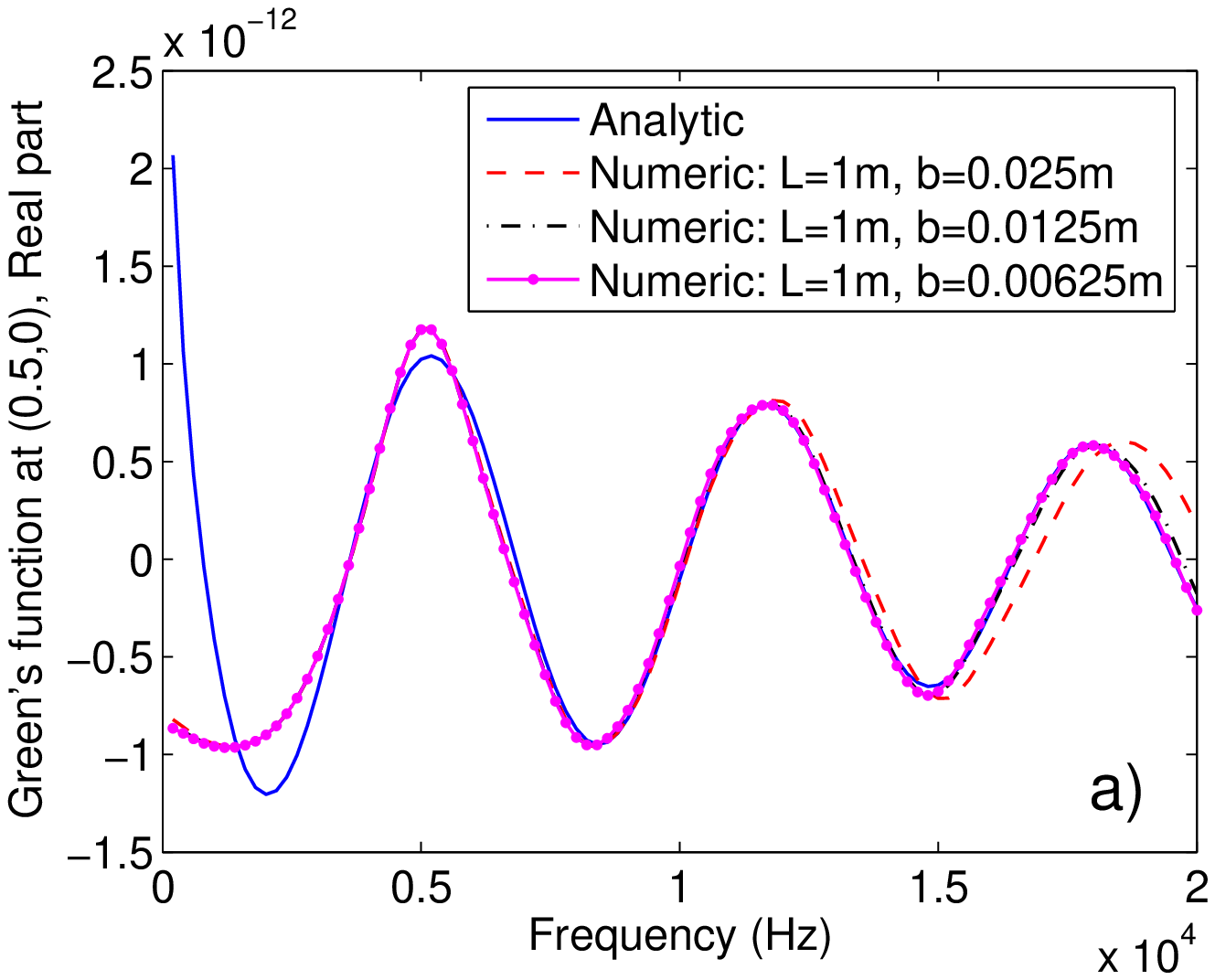,scale=.45}
    \epsfig{file=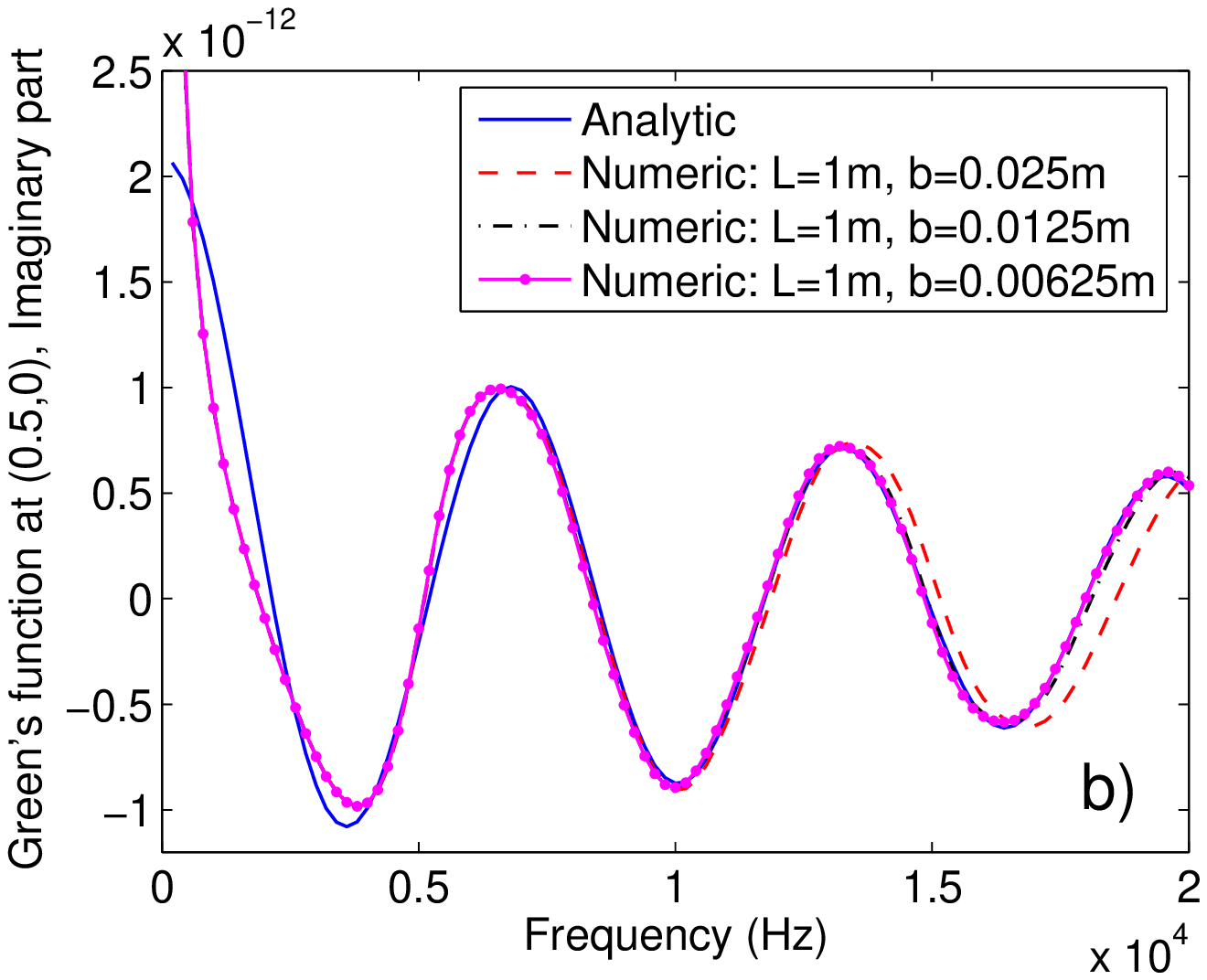,scale=.45}\vspace{0.5cm}
    \epsfig{file=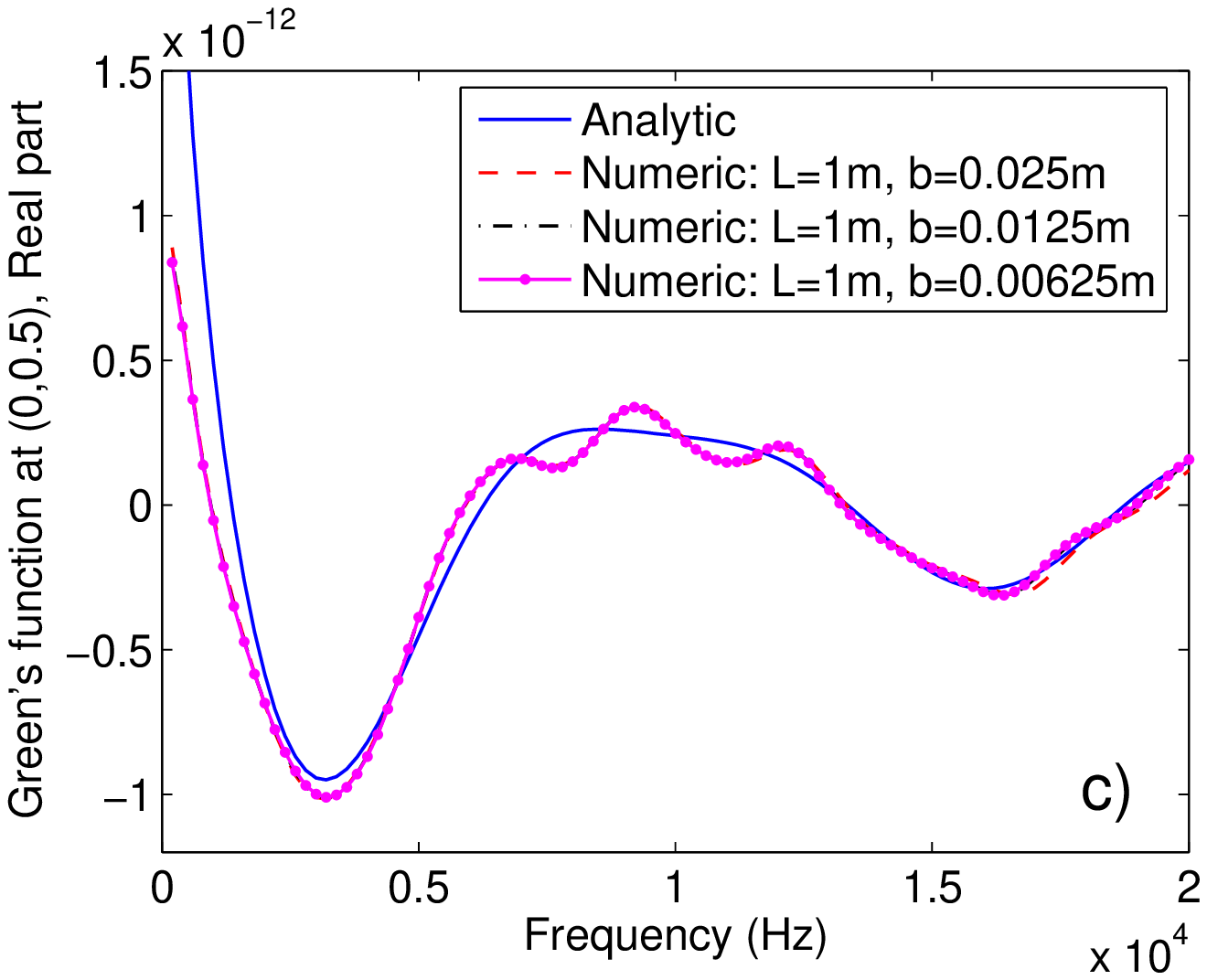,scale=.45}
    \epsfig{file=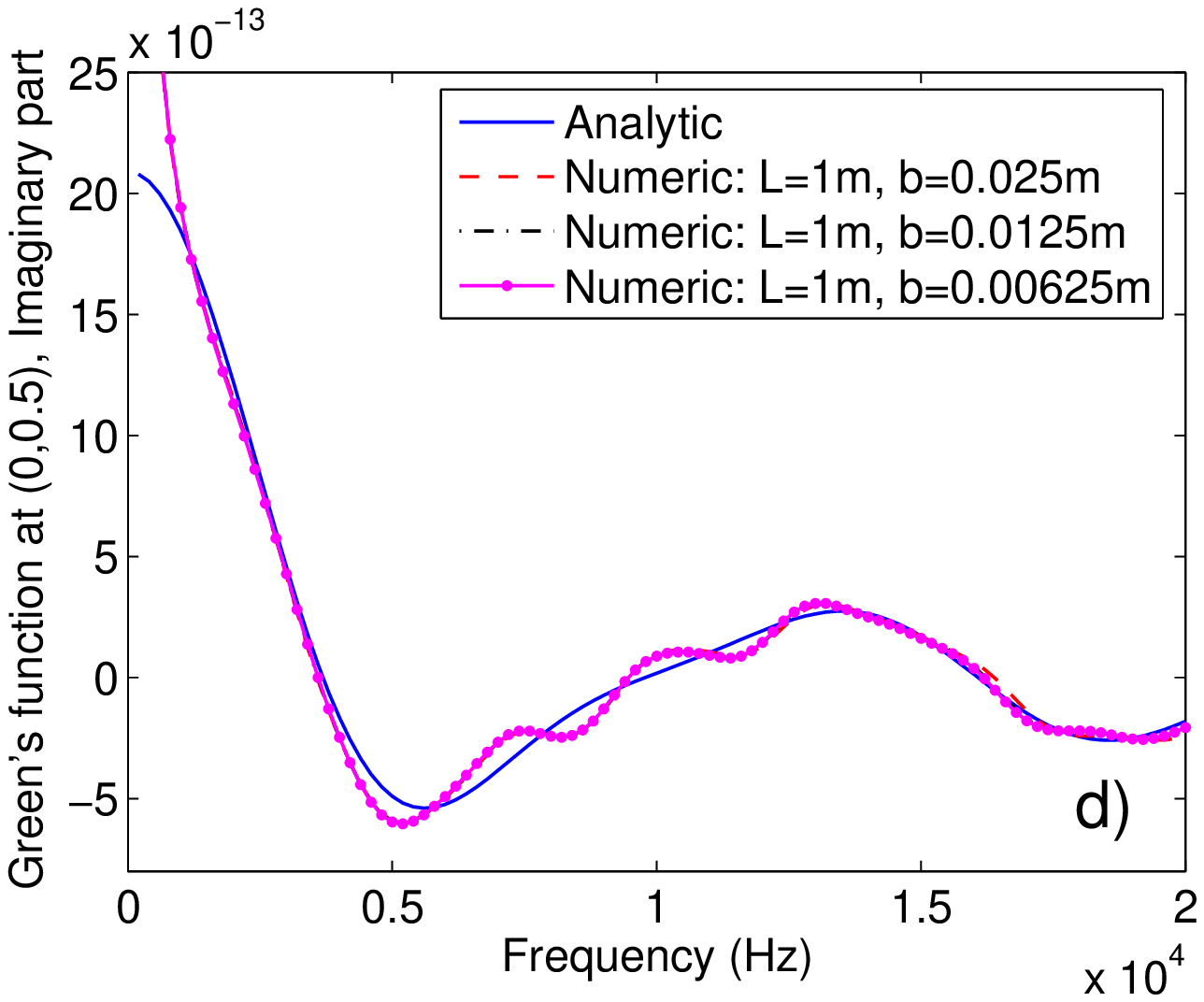,scale=.45}
    \caption{Comparison of analytical and numerical Green's functions
       for the 2D elasticity with different sizes of elements:
      a) Real part at $(0,0.5)$, b) Imaginary part at
      $(0,0.5)$, c) Real part at $(0.5,0)$, d) Imaginary part at
      $(0.5,0)$.\label{fig15}}
  \end{center}
\end{figure}

In figure \ref{fig15}, numerical solutions are compared with
analytical solutions for the point $(0.5,\,0)$ and $(0,\,0.5)$ for
different sizes of the cell. The curves represent the real
and imaginary parts of the first component of the displacement
for an excitation at origin in direction 1. The same
remarks as the previous examples can be made: $1)$ a denser mesh leads to
lower errors over the high frequency band $[8.10^3Hz,\,20.10^3Hz]$; $2)$ for the
low frequency band $[0,\,8.10^3Hz]$, numerical results are
different from the analytical solutions due to the finite size of the
domain $(L=1m)$.

\begin{figure}
  \begin{center}
    \epsfig{file=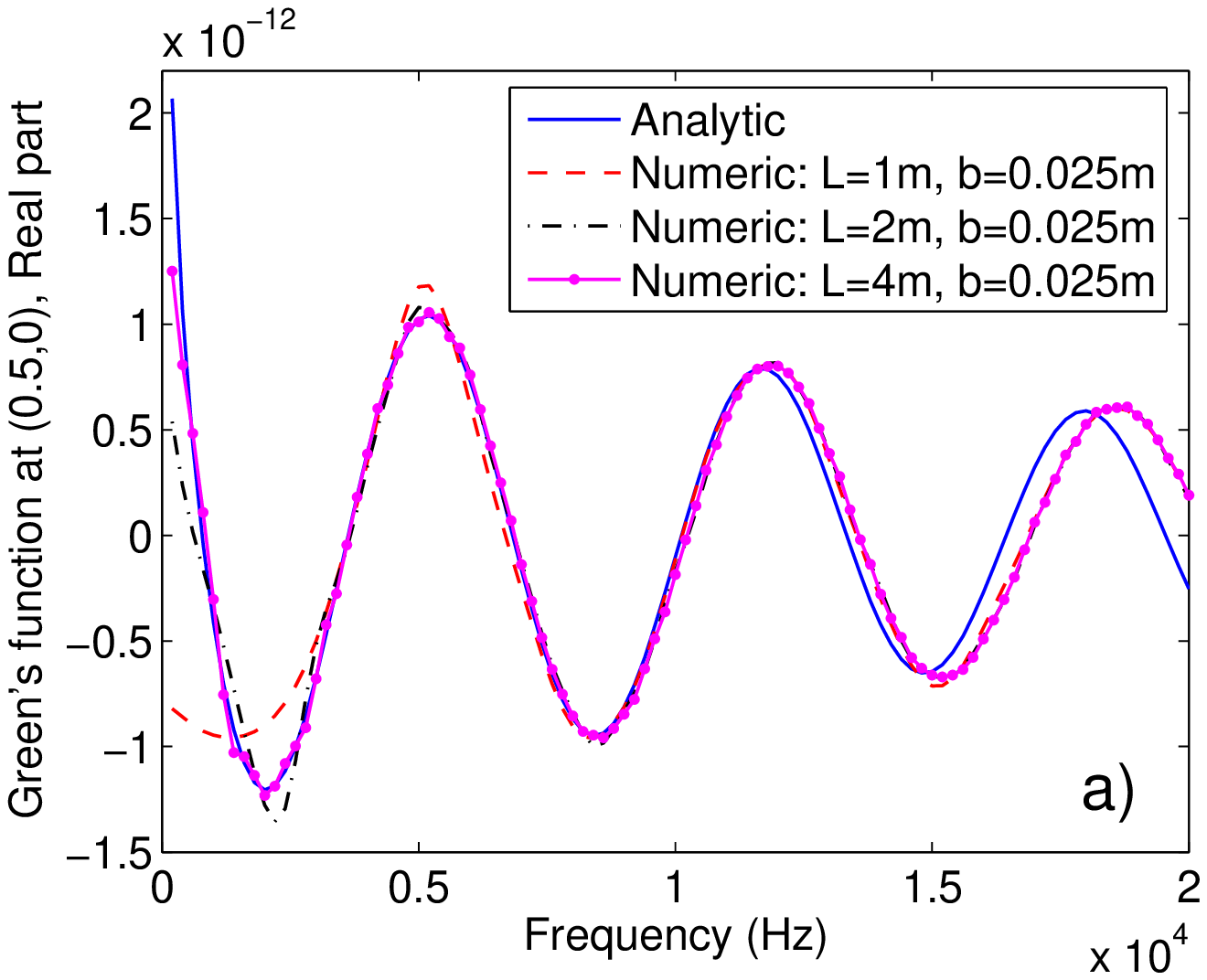,scale=.45}
    \epsfig{file=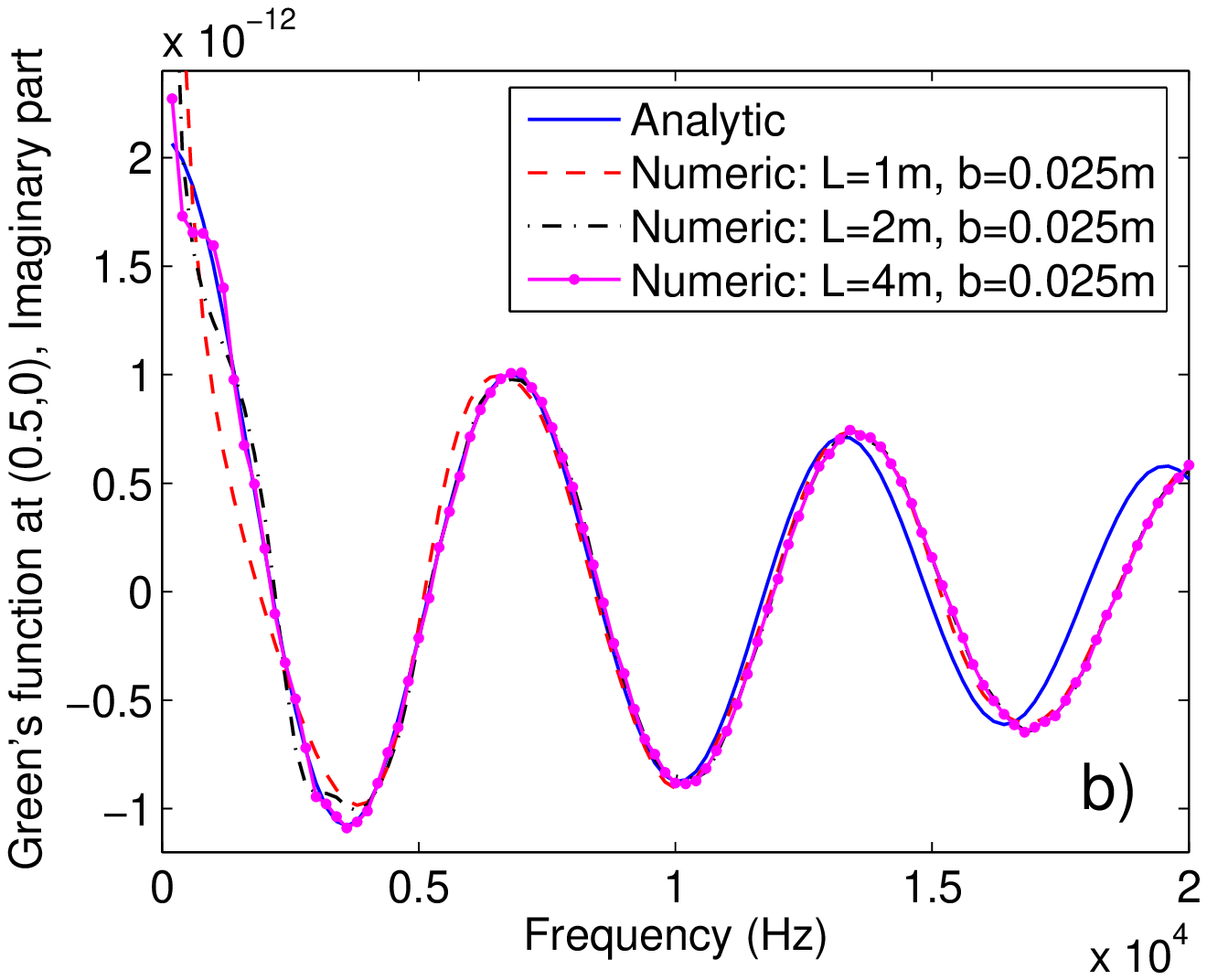,scale=.45}\vspace{0.5cm}
    \epsfig{file=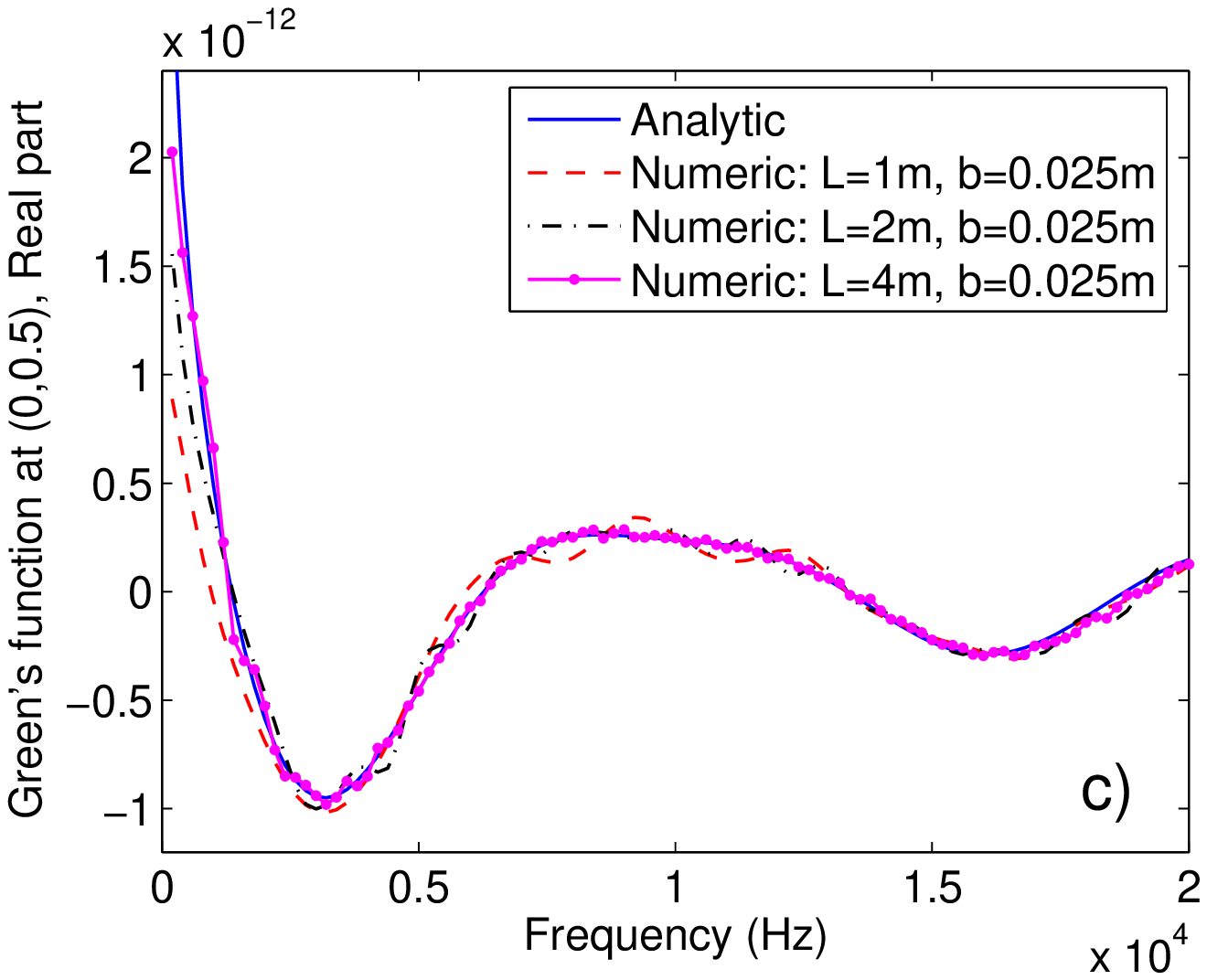,scale=.45}
    \epsfig{file=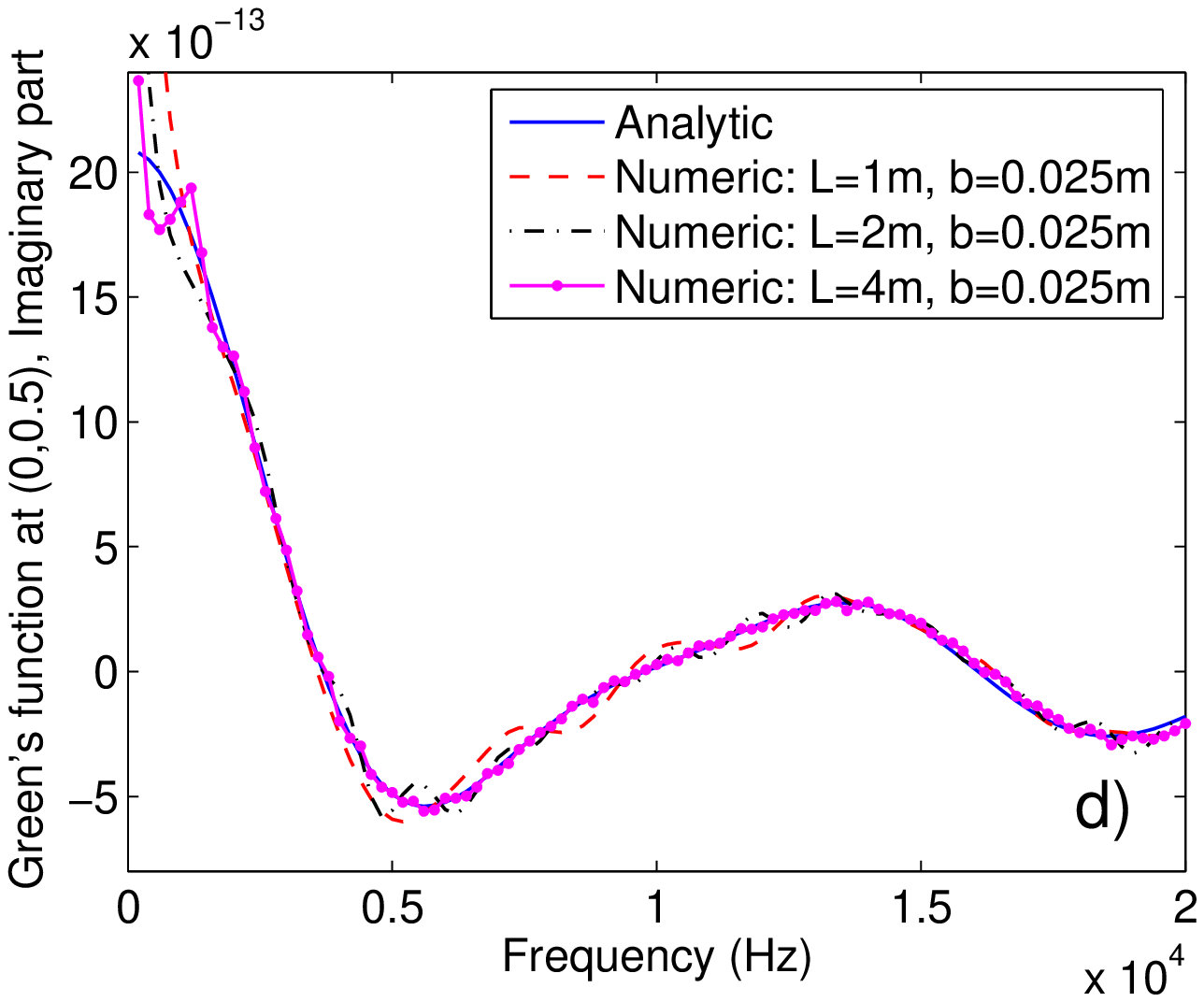,scale=.45}
    \caption{Comparison of analytical and numerical Green's functions
       for the 2D elasticity with different sizes of the domain:
      a) Real part at $(0,0.5)$, b) Imaginary part at
      $(0,0.5)$, c) Real part at $(0.5,0)$, d) Imaginary part at
      $(0.5,0)$.\label{fig16}}
  \end{center}
\end{figure}

In figure \ref{fig16}, the results for these
two points are presented when the size of the domain is increased successively with
$L=1m,\,2m$ and $4m$. In this case, when the size of the cell is
fixed to $b=0.025m$, a larger domain leads to lower errors over the
low frequency band $[0,\,8.10^3Hz]$.
The results are the same as previously in the high frequency band because
in this domain the precision depends on the size of the elements
and not on the size of the global domain.
Some improvements are however seen for intermediate frequencies.

\begin{figure}
  \begin{center}
    \epsfig{file=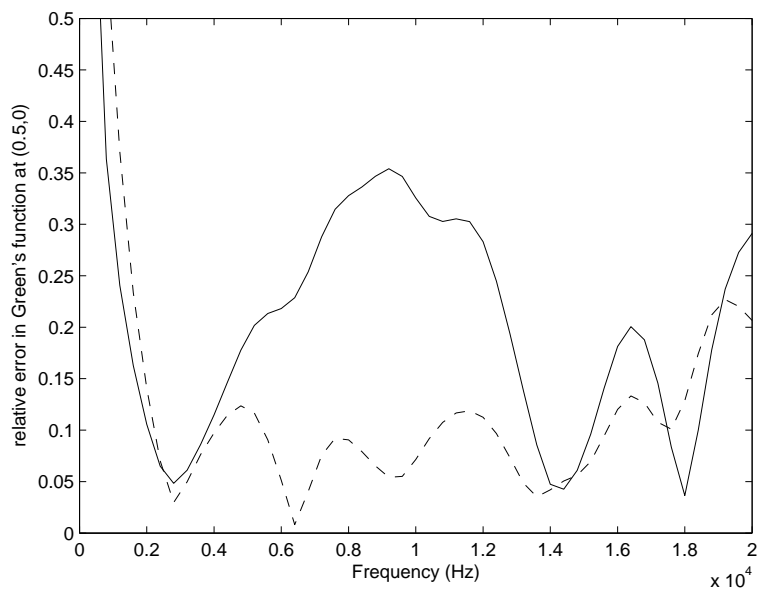,scale=.7}
    \caption{Comparison of relative errors for 0 order $\line(1,0){15}$ and
      second order $--$ boundary conditions for the 2D elastodynamics
      at point $(0.5,0)$ with $L=1m$ and $b=0.025m$.\label{fig17}}
  \end{center}
\end{figure}

In figure \ref{fig17} the error for boundary conditions of zero and second orders
are plotted versus the frequency.
The second order condition is much more accurate for intermediate frequencies
as for the acoustic case.

\section{Conclusion}

In this paper, a method to determine absorbing boundary conditions for
two dimensional periodic media has been presented.
It works directly on the discretized equations.
The boundary condition is first obtained as a global impedance relation
and is then localized into boundary conditions of various orders.
In two examples, good agreements were observed when compared with
analytical solutions.

In any case, the proposed method is efficient because it requires only
the discrete dynamic matrices which can be obtained by any standard
finite element software.
This method could be used for media with more complex behaviors
than those presented in the precedent examples.

\newpage
\clearpage

\newpage
\clearpage

\newpage
\clearpage

\newpage
\clearpage

\newpage
\clearpage

\newpage
\clearpage

\newpage
\clearpage

\newpage
\clearpage

\newpage
\clearpage

\newpage
\clearpage

\newpage
\clearpage

\newpage
\clearpage

\newpage
\clearpage

\newpage
\clearpage

\newpage
\clearpage

\newpage
\clearpage

\newpage
\clearpage

\end{document}